\documentclass[3p]{elsarticle}

\usepackage{hyperref}

\journal{Elsevier}

\usepackage[ansinew]{inputenc}
\usepackage{array}
\usepackage{makecell}
\usepackage{amsthm}
\usepackage{amsfonts}
\usepackage{color}
\usepackage{soul}
\usepackage{booktabs}
\usepackage{rotating}
\usepackage{xcolor,colortbl}
\usepackage{subfiles}
\usepackage{multicol}
\usepackage{stfloats}
\usepackage{url}
\usepackage{ulem}
\definecolor{Gray}{gray}{0.9}
\usepackage{balance}
\usepackage{xcolor}
\usepackage{amsmath}
\usepackage{tikz}
\usetikzlibrary{shapes}
\newcolumntype{C}[1]{>{\centering\let\newline\\\arraybackslash\hspace{0pt}}m{#1}}
\usepackage{rotating}
\usepackage{pdflscape}
\usepackage{afterpage}
\usepackage{capt-of}
\usepackage{longtable}
\usepackage{stfloats}
\usepackage{amsmath,amssymb,amsfonts}
\usepackage{algorithmic}
\usepackage{graphicx}
\usepackage{caption}
\usepackage{textcomp}
\usepackage{longtable}
\usepackage{enumitem}
\usepackage{rotating}
\usepackage{subfigure}
\usepackage{hyperref}
\usepackage{amsxtra}
\usepackage{natbib}
\usepackage{amssymb}
\usepackage{subfigure}
\usepackage{latexsym}
\usepackage{caption}
\usepackage{amsmath,amsfonts,amssymb,amsthm}
\usepackage[ruled,vlined]{algorithm2e}
\usepackage{multirow}
\definecolor{Gray}{gray}{0.9}
\usepackage{balance}
\usepackage{etoolbox}

\makeatletter
\def\ps@pprintTitle{%
   \let\@oddhead\@empty
   \let\@evenhead\@empty
   \def\@oddfoot{\reset@font\hfil\thepage\hfil}
   \let\@evenfoot\@oddfoot
}
\makeatother

\begin{document}
\begin{frontmatter}

\title{Spatial Computing: Concept, Applications,
Challenges and Future Directions}


\author[1]{Gokul Yenduri}
\author[2]{Ramalingam M}
\author[2]{Praveen Kumar Reddy Maddikunta}
\author[3,4,5] {Thippa Reddy Gadekallu*}
\author[6]{Rutvij H Jhaveri}
\author[7]{Ajay Bandi}
\author[8]{Junxin Chen}
\author[9]{Wei Wang}
\author[10]{Adarsh Arunkumar Shirawalmath}
\author[10]{Raghav Ravishankar}
\author[11]{Weizheng Wang}

\address[1]{School of Computer Science and Engineering, VIT-AP University, Amaravati, Andhra Pradesh-522237, India.(email: yenduri.gokul@gmail.com)}

\address[2]{School of Computer Science Engineering and Information Systems, Vellore Institute of Technology, TamilNadu-632014, India.\\
(emails: ramalingam.m@vit.ac.in, praveenkumarreddy@vit.ac.in)}

\address[3]{Zhongda Group, Haiyan County, Jiaxing City, Zhejiang Province, China, 314312.}

\address[4]{College of Information Science and Engineering, Jiaxing University, Jiaxing 314001, China.}

\address[5]{Division of Research and Development, Lovely Professional University, Phagwara, India.\\
(email: thippareddy@ieee.org)}


\address[6]{Department of Computer Science and Engineering, School of Technology, Pandit Deendayal Energy University, Gandhinagar, India. (email: rutvij.jhaveri@sot.pdpu.ac.in)}

\address[7]{School of Computer Science and Information Systems, Northwest Missouri State University, 800 University Dr, Maryville, MO 64468, United States. (email: ajay@nwmissouri.edu)}

\address[8]{School of Software, Dalian University of Technology, Dalian 116621, China. (email: junxinchen@ieee.org)}

\address[9]{Guangdong-Hong Kong-Macao Joint Laboratory for Emotional Intelligence and Pervasive Computing, Artificial Intelligence Research Institute, Shenzhen MSU-BIT University, Shenzhen, 518172, Guangdong, China, School of Medical Technology, Beijing Institute of Technology, Beijing 100081, China (e-mail: ehomewang@ieee.org)}

\address[10]{School of Computer Science and Engineering, Vellore Institute of Technology, TamilNadu-632014, India (e-mail:adarsh.arunkumar2022@vitstudent.ac.in, raghav.ravishankar2022@vitstudent.ac.in)}

\address[11]{Department of Computer Science, City University of Hong Kong, Hong Kong SAR, China (e-mail:weizheng.wang@ieee.org)}


\begin{abstract}
Spatial computing is a technological advancement that facilitates the seamless integration of devices into the physical environment, resulting in a more natural and intuitive digital world user experience. Spatial computing has the potential to become a significant advancement in the field of computing. From GPS and location-based services to healthcare, spatial computing technologies have influenced and improved our interactions with the digital world. The use of spatial computing in creating interactive digital environments has become increasingly popular and effective. This is explained by its increasing significance among researchers and industrial organisations, which motivated us to conduct this review. This review provides a detailed overview of spatial computing, including its enabling technologies and its impact on various applications. Projects related to spatial computing are also discussed. In this review, we also explored the potential challenges and limitations of spatial computing. Furthermore, we discuss potential solutions and future directions. Overall, this paper aims to provide a comprehensive understanding of spatial computing, its enabling technologies, their impact on various applications, emerging challenges, and potential solutions.
\end{abstract}
\medskip
\end{frontmatter}

\begin{table}[t]
\fontsize{8}{8}\selectfont
    \renewcommand{\arraystretch}{1.3}
    \caption{List of key Acronyms.}
    \centering
    \begin{tabular}{|p{1.25cm}|p{5.65cm}|}
        \hline
        \textbf{Acronyms} & \textbf{Description}\\
        
        \hline
        {AI} & Artificial Intelligence \\ 
        \hline
        {AR} & Augmented Reality 
        \\ 
        \hline
        {VR} & Virtual Reality 
        \\
        \hline
        {MR} & Mixed Reality
        \\ 
        \hline
        {BC} & Blockchain
        \\ 
        \hline
        {IoT} & Internet of Things \\ 
       
        \hline
        {GPU} & Graphics Processing Unit \\ \hline
        {AV} & Autonoumous Vehicles 
        \\ 
       
        \hline
        {ML} & Machine Learning 
        \\ 
        \hline
        {6G} & Sixth-Generation \\ 
        
         \hline
        {5G} & Fifth-Generation \\ 
        
        \hline
        {UI} & User Interface \\ 
         \hline
        {GPS} & Global Positioning System \\ 
        \hline
        {HCI} & Human Computer Interface \\ 
        \hline
        {HMD} & Head Mounted Display \\ 
        \hline
    \end{tabular}
    \label{Tab:acronym}
\end{table} 
\section{Introduction} 

\begin{figure*}[h!]
	\centering
\includegraphics[width=0.90\linewidth]{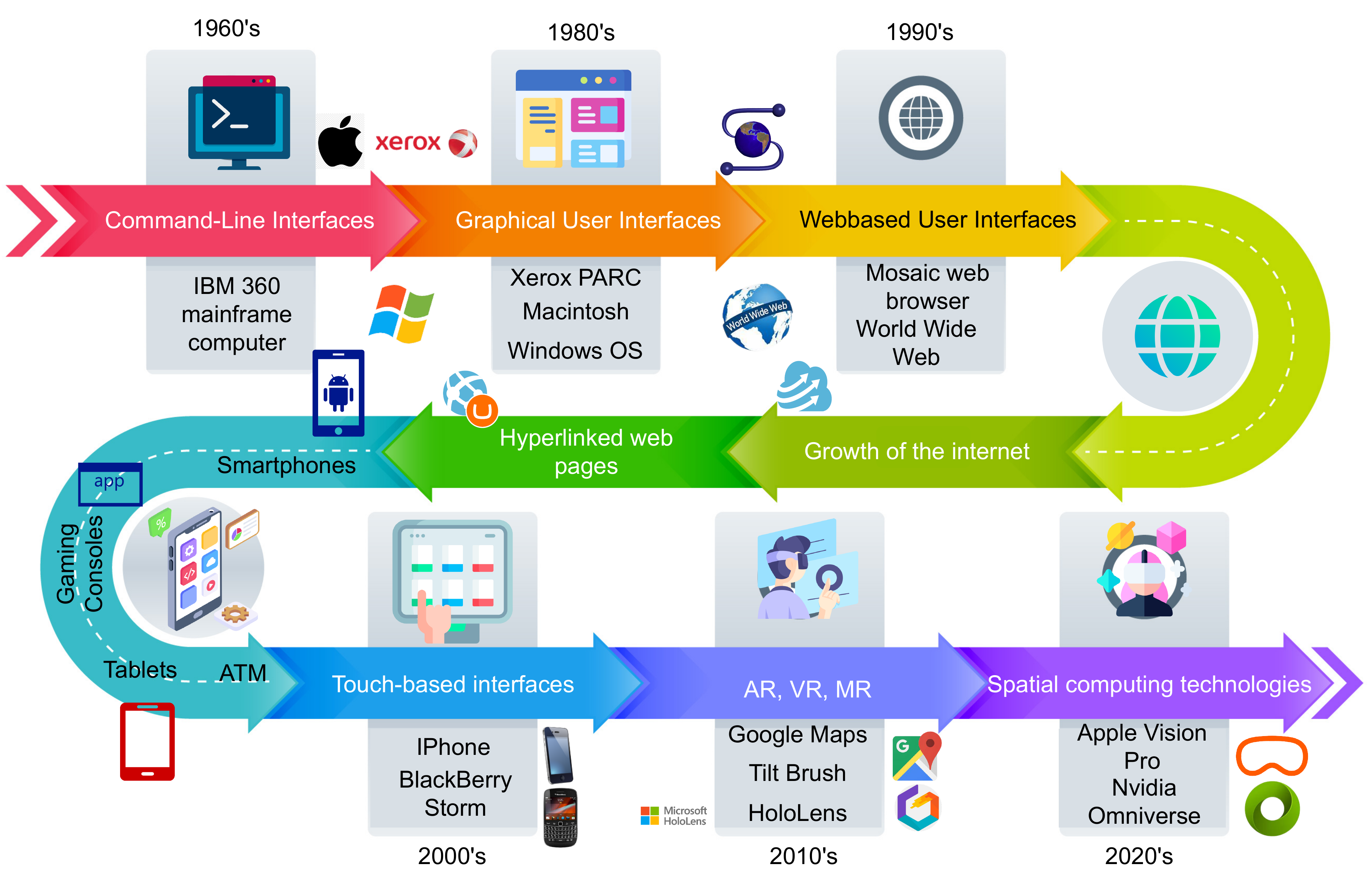}
	\caption{Road Map of User Interfaces.}
	\label{fig:origin}
\end{figure*}

Over the years, User Interfaces (UIs) have seen constant development and improvement. Command-line interfaces, which have been around since the 1960s, require users to type in commands in order to interact with computers. In the 1980s, graphical user interfaces were developed to make computers more user-friendly. These interfaces used windows and icons to represent commands and data. In the 1990s, as the internet grew in popularity, web-based interfaces replaced traditional methods. Then, in the middle of the 2000s, smartphones and tablets appeared, leading to the era of touch-based interfaces and radically changing the ways in which we engage with electronic gadgets. The road map of UIs development over the years is depicted in Fig. \ref{fig:origin}.

The advancement of modern technology like AI, the IoT, digital twins, AR, VR, and MR is now leading us to a new generation of UIs \cite{lee2021all}. These modern technologies are blurring the distinction between real and virtual environments. This seamless and highly efficient blending of the physical environment with the digital realm creates endless possibilities for highly immersive, interactive experiences. Spatial computing is one such technology that encompasses many enabling technologies, including the IoT, digital twins, ambient computing, AR, VR, and AI. The term spatial computing was first coined by Simon Greenwold in 2003, whereby it refers to the interaction between humans and machines, in which a machine has the ability to retain and manipulate referents to real-world objects and surroundings \cite{greenwold2003spatial}. 

Many digital technologies, such as MR, AR, and VR, may facilitate virtual immersion. These technologies may be summed up as XR and belong to the immersive technologies category. The future of "Spatial Computing" will be significantly influenced by XR technologies. Spatial computing technologies enable us to transition from an existing global data network with flat or restricted dimensions to an emerging immersive global data ecosystem that improves performance and senses by giving virtual features to actual things and vice versa. Creating connections between data and representations in larger physical dimensions requires creating a new, complex, multimodal user interaction paradigm as well as the supporting infrastructure. As a result of the paradigm shift from representational to localised and often embodied content creation and consumption, bodily awareness and self-presence are created within the material and the information it conveys. This is seen in an entirely different way as a hybrid or virtual synthetic human experience \cite{mystakidis2023immersive}. In 2023, spatial computing gained traction because of Apple's Vision Pro, a headgear with AR, VR, and MR capabilities \cite{waisberg2023apple}. Spatial computing devices map the surroundings of the user by using data from cameras and other sensors. To determine the geometry of the objects surrounding the user, this data is often passed through one or more algorithms. The classification of the items in the area of vision may also involve image recognition on more sophisticated devices. The technology can superimpose virtual objects that integrate with the real world once it has a spatial map of the area and an understanding of the objects inside.

\subsection{Related works and Contributions}
Spatial computing had a market value of around \$102.1 billion in the year 2021. Projections indicate that this industry is expected to expand to around \$544.6 billion by the year 2032 \cite{futuremarketinsights}. Spatial computing-related studies are widely available and have increased over the past few years. The evident potential of spatial computing has been shown to have a wide range of applications, which include education, healthcare, tourism, the military, navigation, gaming and entertainment, architecture and design, and e-commerce. 

Akers et al. examined the use of spatial computing in the context of a remote classroom in a case study. A 10-week undergraduate course was administered exclusively online, using a combination of traditional teleconferencing software and MR spatial computing technology (specifically, Magic Leap One headsets). Additionally, an avatar-mediated social engagement programme was included in the course structure. The course concluded with a virtual poster session in which project outcomes were presented using spatial MR. An initial study was conducted on the experiences of the students, using questionnaires and interviews as methods of study. The students expressed their satisfaction with the use of spatial computing during the poster session, noting that they saw it as having advantages compared to 2D video conferencing. The study emphasised many notable benefits, including enhanced opportunities for social interaction with peers and instructors, heightened perceptiveness in understanding others' points of interest, and better efficiency in sharing and collaborating on three-dimensional project content \cite{akers2020mixed}.

The use of smartphones and GPS-enabled devices presents new possibilities for the integration of spatial computing applications. The use of sensors on these devices can facilitate the integration of geoinformatics within the domains of social media and disaster management. The use of real-time event data acquired from on-site responders and emergency management communities has the potential to enhance situational awareness, enable informed decision-making, optimise resource allocation, and facilitate enhanced disaster response. In their study, Adam et al. conducted a thorough examination of the capabilities and effectiveness of one such initiative, the Social Media Alert and Response to Threats to Citizens (SMART-C) initiative. This initiative is administered by the Science and Technology Directorate (DHS-S\&T) of the United States Department of Homeland Security. The authors also investigated the unique and specific challenges that emerge from the fusion of social media and spatial computing in the field of disaster management. The main goal of this programme is to provide citizen participatory sensing capabilities that may be used for decision support in all phases of a disaster, using a wide array of devices and modalities \cite{adam2012spatial}.

Mouayad Masalkhi et al. conducted a study on the potential impact of the Apple XR headset, a spatial computing platform, on improving accessibility for individuals with visual impairments. Due to its exceptional 4-K displays per eye and 5000 nits of brightness, this particular headgear can enhance the visual experience and expand opportunities for those with visual impairments, according to the authors. The authors conducted an in-depth examination of the technical specifications and factors relating to accessibility, providing significant insight into the potential benefits of this technology for those with visual impairments. In addition, they claimed that this technology would play an important role in the fields of ophthalmology, surgical training, diagnostics, and educational purposes. The authors also suggested the importance of ensuring test-retest reliability for the clinical application of this technology. The authors came to the conclusion that spatial computing technology has the potential to improve the well-being of people with visual impairments \cite{masalkhi2023apple}.

An experiment conducted by Soyoung Jung et al. aimed to evaluate the impact of spatial embodiment in augmented reality on medical attitudes towards the self. In this study, the presence was shown through a variety of devices, such as a two-dimensional screen, a three-dimensional dummy, and the participant's own bodies. A total of 90 college students took part in the experiment. They watched advertisements that showed virtual pictures of foetuses and X-rays of the lungs. The results of the study showed that public service announcements with increasingly detailed physical interfaces improve the feeling of "being there," which is also called "spatial presence." This evoked adverse emotions towards smoking. When the dual model process is mediated by the spatial presence, it raises adverse attitudes towards smoking that increase the behavioural intention of participants to engage with the campaign \cite{jung2019augmented}.

In order to investigate how ambient temperature and virtual tourism audio-visual settings affect tourists' thermal perception and comfort level, Xiao-Ting Huang et al. carried out an experimental laboratory investigation. A $3\times2\times2$ experimental design was used in the investigation, and 180 individuals made up the participant group. To build virtual tourism destinations, VR equipment and microclimate simulation technologies were used. Additionally, electrocardiogram recording devices were used to evaluate each person's physiological markers. The study's findings imply that participants' thermal perception and certain physiological markers are significantly impacted by virtual tourism spatial settings, which include ambient temperature and audio-visual conditions. The participants' comfort level with respect to temperature is also greatly influenced by these spatial conditions. The research also showed that physiological reactions serve as intermediaries, transferring the impact of tourist spatial circumstances to thermal comfort. This work makes a substantial contribution to the corpus of knowledge on virtual tourist experiences and geographic circumstances in the Metaverse by using spatial computing. It also offered theoretical and managerial recommendations for the creation of immersive virtual tourism environments \cite{huang2023experimental}.

The above recent works illustrate the significant impact of spatial computing and its significance across various sectors. Therefore, it is important to conduct an in-depth survey on spatial computing in order to get a thorough understanding of its underlying concept, various applications, existing challenges, and potential future prospects.

The above observations have motivated us to conduct this comprehensive review of spatial computation.
Principal contributions of this study include:
\begin{itemize}
\item Potential spatial computing applications are presented.
\item  Some of the most important spatial computing-related research and industry projects are discussed.
\item Several challenges associated with the integration of spatial computing with diverse applications are discussed. In addition, we also provided the possibilities for future research that encourage researchers and industry to pursue further investigation on spatial computing.
\end{itemize}

\subsection{Systematic Literature Survey}
This systematic literature survey presents a comprehensive examination of spatial computing, using a rigorous analysis of scholarly literature from several reliable sources. Our research largely concentrated on scholarly publications that undergo a peer-review process, as well as reputable national and international conferences, seminars, books, symposiums, and journals known for their high-quality content. In order to establish the credibility and legitimacy of our sources, we investigated reputable archives such as Google Scholar and arXiv, as well as publications from prominent databases like IEEE, Springer, Elsevier, Taylor \& Francis, and Wiley, among others. In order to identify appropriate sources and publications in the field of spatial computing, we employed specific keywords, including spatial computing, spatial mapping, spatial awareness, spatial interaction, spatial user interface, spatial computing applications, spatial computing devices, spatial computing development platforms, spatial computing content creation, spatial computing user experience, and spatial computing challenges and ethics. Subsequently, a thorough examination of all the obtained articles was conducted, whereby the screening process included evaluating the titles and removing any papers that had substandard content. Subsequently, the abstracts of the remaining publications were examined in order to ascertain their respective contributions. During the final phase of our literature study, we systematically retrieved the essential data required for our analysis. By following these stages, we assured that our study was based on sources of high quality and credibility \cite{kitchenham2009systematic}.

\subsection{Paper Organization}
The rest of the paper is organized as follows: Section 2 presents the preliminaries of spatial computing and the motivation of the work. Section 3 discusses the potential applications of spatial computing. In Section 4, we highlighted some of the exciting projects that are related to spatial computing. Section 5 includes open issues, challenges, and future research directions in spatial computing. Finally, we conclude the paper in Section 6 by summarising the key findings and contributions of this study.

\section{Preliminaries}
Spatial computing is the use of technology that integrates digital information and virtual content into the physical world, enabling users to interact with and perceive digital content in a spatial context. Digital content refers to any information related to virtual objects, animations, holograms, etc. Within a given space, spatial context involves understanding and interpreting virtual objects, entities, and events with respect to their physical location and the geometric relationships between them. By utilizing these relationships, spatial computing ensures that virtual objects are placed correctly to interact with real objects and respond accurately to user movements and actions. 

In traditional HCI experiences, users interact with the user interfaces of websites and mobile devices. The goal of spatial computing is to enhance HCI by providing intuitive ways of engaging with digital content as if it were part of their physical environment \cite{clifton2016design}.

AR: AR involves overlaying digital information, such as images, videos, or 3D models, onto the real world. AR enhances our perception of reality by adding virtual elements that interact with and augment our physical environment. AR can be experienced through devices like smartphones, tablets, smart glasses, or heads-up displays. It allows users to see and interact with virtual content while still being aware of the real world \cite{xiong2021augmented}.

VR: VR aims to create a fully immersive digital experience by simulating a virtual environment that users can interact with using specialized hardware, such as VR headsets. VR completely replaces the user's real-world surroundings and transports them to a simulated, computer-generated environment. Users perceive themselves as being present in a different world and can explore and interact with virtual objects and spaces \cite{scavarelli2021virtual}.

Spatial Computing: Spatial computing is a broader concept that encompasses both AR and VR but goes beyond them. It refers to the use of technology to understand and interact with the physical world in a spatial context. Spatial computing combines the real and virtual worlds, allowing users to perceive and interact with digital content within their physical environment. It involves technologies like computer vision, sensors, tracking systems, and spatial mapping to create a seamless integration of digital and physical elements \cite{mohamed2023deep}.

\begin{table*}[]
\caption{Immersive Technolgoies Comparison Matrix.}
\label{tab:comparison_table}
\centering
\resizebox{\textwidth}{!}{%
\begin{tabular}{ | m{2.58 cm} | m{4.5 cm}| m{4.5 cm} | m{5 cm}|}
\hline
\textbf{Parameters} & \textbf{AR}                                               & \textbf{VR}                                             & \textbf{Spatial Computing}                                                                   \\ \hline
Definition          & Overlays virtual content onto the real world, enhancing perception of reality & Creates a fully immersive simulated environment, replacing the real world & Blends the real and virtual worlds, enabling interaction and perception in a spatial context \\ \hline
Hardware            & Smartphones, tablets, AR glasses, heads-up displays                           & VR headsets, HMDs, immersive devices                                      & AR glasses, VR headsets, smart devices                                                       \\ \hline
Interaction         & Gesture-based, touch-based, voice commands                                    & Controller-based, hand tracking                                           & Gesture-based, touch-based, voice commands, tracking systems                                 \\ \hline
Immersion           & Partial immersion, overlaying virtual content on the real world               & Full immersion, replacing the real world with a virtual environment       & Varies depending on the application and hardware used                                        \\ \hline
Scope               & Enhancing the real world with virtual content                                 & Creating immersive virtual experiences                                    & Enabling integration of real and virtual worlds in a spatial context                         \\ \hline
Key Technologies    & Computer vision, sensors, overlay techniques                                  & Headsets, display technologies, tracking systems                          & Computer vision, sensors, tracking systems, spatial mapping                                  \\ \hline
Use Cases           & Gaming, education, training, navigation, entertainment                        & Gaming, simulations, training, entertainment                              & Architecture, interior design, industrial applications, collaborative workspaces             \\ \hline
Examples            & Pokémon Go, Snapchat filters                                                  & Oculus Rift, HTC Vive                                                     & Microsoft HoloLens, Magic Leap                                                               \\ \hline
\end{tabular}
}
\end{table*}

To implement this, search and read a few recent papers on spatial computing. Here are some of the references. Memory \cite{lundqvist2023working}, decision support in biomedical area \cite{farhadloo2023spatial}, robotic embodiment \cite{lobov2023spatial}, devices \cite{savio2023accelerating}, future thing is spatial computing \cite{garg2023spatial}, machine vision with spatio-temporal photonic computing \cite{zhou2023ultrafast}, big data analysis \cite{gao2023research}, spatial digital twin \cite{ali2023enabling}, digital twin \cite{2022}. Table \ref{tab:comparison_table} provides the comparison of AR, VR and spatial computing.

\subsection{Enabling Technologies of Spatial Computing}
\begin{figure*}[h!]
	\centering
\includegraphics[width=0.90\linewidth]{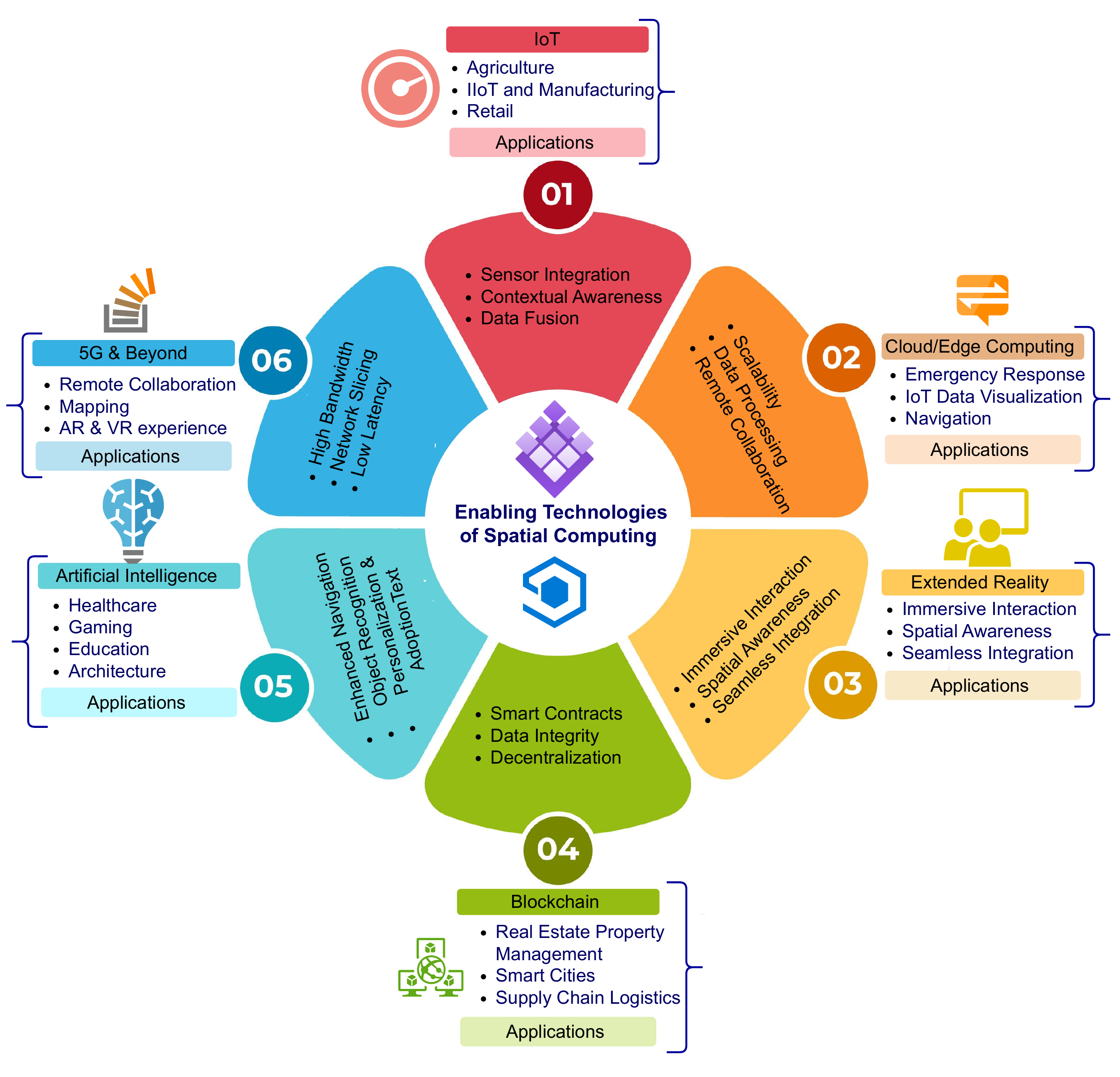}
	\caption{Enabling Technologies of Spatial Computing.}
	\label{fig:enabling}
\end{figure*}

This section breaks down the essential technologies that make spatial computing possible. These are the technologies that merge the digital and real worlds seamlessly, creating immersive experiences. Fig.\ref{fig:enabling} depicts the enabling technologies and their applications that can contribute to spatial computing.
\subsubsection{Artificial intelligence}
To combine the physical environment and digital information of virtual objects artificial intelligence is essential to understand the scene, identify and track objects in the virtual environment, recognize the gestures, detect the interaction between objects, understand and handle object occlusion, and map them with the real-time spatial layout. Using mixed reality headset equipped with cameras and sensors, one can interact with their surroundings in spatial computing. This experience is powered by AI, which combines the physical environment with digital information of virtual objects. Computer vision, an essential technology within AI, plays a vital role in analyzing the scene by identifying objects like walls and furniture and creating a 3D representation of the environment. The device recognizes and tracks virtual objects through computer vision, allowing users to manipulate them naturally using gestures. Additionally, the device ensures that virtual objects appear realistically behind real-world objects by handling occlusions. Moreover, computer vision helps update its spatial understanding as you move around for accurate placement and persistence of virtual objects. These technologies create an immersive experience where physical and virtual worlds coexist seamlessly in gaming, design, training, education, etc \cite{lv2022blocknet}.

Tracking systems have a critical role in spatial computing, facilitating smooth user interactions with digital content within the physical environment \cite{plopski2023tracking}. These systems are necessary for precisely positioning virtual objects, interpreting user actions, and ensuring an immersive experience. Inside-out tracking \cite{whitmire2019aura} is commonly used in VR headsets and AR glasses to monitor the user's movement and surroundings through integrated cameras and sensors. External cameras or sensors positioned around the room are employed in outside-in tracking \cite{radle2018polartrack} to accurately track markers on the user or their spatial computing device. Hand and gesture recognition systems utilize cameras or sensors to identify various hand poses and gestures, allowing natural interaction with virtual objects. Body tracking captures full body movements, while eye tracking monitors gaze direction, enhancing the user experience with dynamic foveated rendering and gaze-based interactions. Spatial computing devices also utilize positional audio monitoring for an immersive auditory experience. Moreover, environmental mapping techniques create real-time maps of the user's environment using  Simultaneous Localization and Mapping (SLAM) techniques that enable virtual content to anchor accurately while maintaining spatial coherence. By combining these tracking systems, spatial computing devices provide an interactive and immersive experience where physical elements coexist seamlessly with virtual ones, making possible a wide range of applications from gaming and entertainment to education, training, etc.

\subsubsection{IoT}
In spatial technology, the IoT plays a critical role by embedding hardware that collects digital information. Spatial computing relies on various sensors and cameras integrated into the physical environment to gather data about the location of virtual objects and analyze their relationships in a spatial context. These sensors may include depth sensors, RGB cameras, infrared sensors, and LiDAR (Light Detection and Ranging) sensors, which capture information about the user's surroundings, providing the system with a deep understanding of the spatial context.

In addition to sensors, spatial computing requires display devices to present virtual content to the user. Cutting-edge display technologies are used for this purpose. HMDs, such as VR headsets or AR smart glasses, offer immersive experiences where digital elements are seamlessly overlaid in the real world. Alternatively, spatial computing can be experienced on devices like smartphones, tablets, or projection systems, depending on the specific application's requirements. To achieve real-time responsiveness, the collected images and data need to be processed promptly. GPUs play a crucial role in spatial computing, especially when rendering high-quality and interactive 3D graphics in real-time. GPUs handle the computational demands of rendering complex scenes, managing lighting and shading effects, and ensuring smooth and realistic visuals for a more immersive experience \cite{mohanto2022integrative}.

Spatial computing, with its integration of hardware, sensors, cameras, and GPUs, allows users to perceive and interact with digital content in a spatial context. This technology facilitates the seamless blending of the real and virtual worlds, enabling users to engage with virtual elements as if they were part of their physical environment. The combination of IoT, sensors, display devices, and powerful GPUs brings spatial computing to life, offering immersive and intuitive experiences that revolutionize how we interact with technology.

\subsubsection{5G and Beyond}
The integration of 5G and beyond technology with spatial computing holds tremendous potential to revolutionize our interactions with digital environments, particularly in areas like gaming and education. By facilitating a deeper connection between physical and digital spaces, 6G enables seamless transitions between virtual reality and the real world. According to Akyildiz et al., \cite{6GAkyildiz2020}, the increased bandwidth, ultra-low latency, enhanced connectivity, and higher data rates of 6G networks empower spatial computing applications to process and display vast amounts of data in real time. The practical applications of spatial computing span diverse industries, including architecture, design, healthcare, and education. With the advanced capabilities of 6G, these applications can be enhanced, allowing for real-time data processing on an unprecedented scale. In spatial computing, the integration of 6G technology facilitates exploring virtual environments and manipulating complex models with minimal lag time, resulting in a truly immersive experience that can drive innovations across various fields.

The combination of 6G and spatial computing paves the way for haptic communication and tactile feedback. Users can experience a higher sense of immersion and engagement, feeling and touching virtual objects as close as the real environment. This not only enriches the user experience but also enhances the sharing of digital information within the virtual environment. The combination of 6G with spatial computing brings a transformative era, offering seamless, responsive, and immersive experiences that redefine how we interact with digital content and the physical world. It opens up possibilities for novel applications, enriching education, entertainment, design, and various other industries, and propelling technology into a new era of interconnectedness and engagement.

\subsubsection{Cloud/Edge Computing}

Spatial computing often relies on both cloud and edge computing as part of a distributed architecture. Cloud computing and edge computing are integral components of the spatial computing landscape, a technology that fuses the digital and physical realms to create immersive and context-aware experiences. Cloud computing, which involves leveraging remote servers on the internet to manage, process, and store data, plays a crucial role in spatial computing. This technology provides critical functions, such as data storage for voluminous 3D models, maps, and geospatial information. Cloud storage solutions offer scalability in spatial applications. Cloud computing supports data processing for spatial applications, including rendering complex 3D graphics and simulations. By offloading resource-intensive processes from edge devices to the cloud, spatial experiences become smoother and more responsive. It also facilitates data sharing and collaboration, allowing multiple users or devices to access and collaborate on the same data or experiences. The cloud-based analytics tools can process and analyze the data generated by spatial computing applications, offering valuable insights for data-driven decision-making \cite{ikegwu2022big}.

In contrast to the centralized approach, edge computing takes a different approach by processing data closer to the source or ``edge'' of the network. Within the context of spatial computing, edge computing has various advantages. Low-latency interactions are a critical requirement for spatial experiences demanding real-time responsiveness. Edge computing minimizes the round-trip time for data to travel to and from the cloud, enhancing applications such as AR/VR, autonomous vehicles, and robotics. Additionally, it prioritizes privacy and security by allowing spatial data to remain local or within a specific geographical region, reducing the risk of data breaches and facilitating compliance with data privacy regulations. Finally, edge devices can perform on-device processing, reducing the computation load on the cloud, saving bandwidth and processing costs, and catering to resource-constrained or cost-sensitive environments. 

\subsubsection{Extended Reality}
XR, encompassing technologies like AR, VR, and MR, plays a vital role in enabling and enhancing spatial computing. AR, overlays digital information onto the real-world environment, contributing to spatial understanding. In the context of spatial computing, AR is used to provide context-aware information about physical objects or locations, enriching the immersive experience. AR also facilitates spatial interaction, allowing users to engage with digital objects in the real world, supporting gesture recognition, touchless interactions, and spatially aware interfaces. AR is invaluable for navigation applications within spatial computing, offering visual cues and directions overlaid on the real world, simplifying complex environmental navigation \cite{templin2022using}.

VR, on the other hand, creates fully immersive, computer-generated environments where users can step into virtual worlds. In spatial computing, VR is harnessed for a multitude of purposes, such as training simulations, virtual tours, and spatial design. It provides users with a detailed sense of presence, making it a potent tool for immersive spatial experiences. Additionally, VR enables virtual collaboration, allowing geographically dispersed individuals to convene in a common virtual space, which is particularly beneficial in fields like architecture, engineering, and remote teamwork. VR's capability for visualizing spatial data and 3D models is also leveraged, enhancing the comprehension and analysis of spatial information.

MR bridges the gap between the physical and digital worlds by seamlessly blending of both elements. In spatial computing, MR is employed to blend digital content into the real world, creating immersive, mixed environments. Spatial anchors are often used in MR to accurately position digital objects within the physical environment, ensuring precise alignment and interaction. MR devices offer hands-free interaction through gaze-based selection and gesture control, enhancing the ease of interaction with spatial content without the need for physical controllers.

\subsubsection{Blockchain}
Spatial computing, the innovative fusion of digital information with the physical world, is transforming the way we interact with and understand our surroundings. To further enhance this immersive experience and address critical challenges, BC technology is emerging as a valuable enabling tool. BC, is known for its secure, decentralized, and transparent nature, is finding applications in various aspects of spatial computing, ensuring the integrity and reliability of geospatial data, enhancing location-based services, and facilitating decentralized applications. The following paragraphs, explore the diverse ways in which BC is harnessed within the realm of spatial computing, from data management and smart contracts to security and interoperability.

BC technology is increasingly utilized in spatial computing to establish a secure and tamper-proof ledger for geospatial data. This application ensures the integrity and authenticity of location-based information, a crucial aspect of spatial computing for various applications like navigation, logistics, and urban planning \cite{kovacova2022algorithmic}. BC also enhances location-based services by providing a decentralized and trustless platform for users to interact with their location data. This empowers users to have more control over their location information, allowing them to share it securely with third-party services while maintaining their privacy. BC, helps in data provenance and traceability, creating an immutable record of data sources and changes, thus ensuring transparency and trust in location-based information.

BC's capabilities extend to managing spatial transactions with the use of smart contracts. These self-executing contracts automate various activities related to spatial data, such as location-based advertising, microtransactions for access to specific spatial data, and geospatial asset trading. BC also supports the creation of decentralized spatial data marketplaces, where users can securely buy, sell, or trade geospatial data and services \cite{serrano2022verification}. The secure and transparent nature of BC is well-suited for these marketplaces, promoting innovation in spatial computing. BC enhances the security of location data through encrypted and decentralized storage solutions, protecting sensitive spatial information from unauthorized access or data manipulation.

BC's potential in spatial computing extends to the development of decentralized Applications (dApps) that leverage its security and transparency. These dApps provide a wide range of services, including decentralized navigation, immersive AR/VR experiences, and location-based gaming, enhancing spatial computing experiences. BC also plays a role in crowdsourced spatial data, incentivizing users to contribute to real-time datasets through token rewards, ensuring the accuracy and reliability of such data. Additionally, it facilitates interoperability among various spatial computing platforms and devices, promoting the seamless integration of spatial data into a multitude of applications \cite{ghosh2022strove}. However, it's essential to consider potential challenges and complexities, such as scalability and energy consumption, associated with BC implementation in spatial computing.

\subsection{Motivation}
In today's rapidly evolving world, requirements in various sectors are escalating at an unprecedented pace. Whether it be healthcare, education, manufacturing, or military, each industry is under enormous pressure to meet the growing needs of a global population. In healthcare, the need for efficient patient care and remote monitoring has rushed for the incorporation of advanced technological solutions. Education has undergone a digital revolution, with the demand for remote learning tools and interactive virtual classrooms. While the military sector relies on cutting-edge technologies for surveillance, communication, and warfare strategies. Architects and urban planners necessitate advanced tools to design sustainable and efficient spaces. The manufacturing sector is adopting automation and smart technologies to improve productivity. Meanwhile, the gaming industry is continually pushing boundaries to provide immersive and realistic experiences. These sectors are experiencing exceptional demands, and meeting them requires innovative technologies with immersive features.
Spatial computing is one such technology with immersive abilities more specifically the fusion of XR technologies. It has the potential to revolutionize various sectors through its unique features. In healthcare, surgeons can use AR to overlap patient data during surgeries, allowing more precise procedures. Education can be valued by immersive VR classrooms that bring students together in a virtual space, making learning more engaging and accessible. The military can make use of AR for enhanced situational awareness, planning, and training simulations. Architects can use VR to create 3D models of buildings, permitting for better design visualization and collaboration. In manufacturing, spatial computing can make more efficient processes through AR-guided assembly and maintenance. Additionally, the gaming industry can make use of this technology to create breathtaking virtual worlds and experiences. Spatial computing has the potential to address the multidimensional demands of these sectors by offering innovative, interactive, and efficient solutions, ultimately shaping the future of each industry.

As we traverse through an era marked by exceptional challenges across diverse industries, the attraction of spatial computing lies in its transformative capabilities. The following will shed light on the key features that make spatial computing a technological revolution in overcoming contemporary challenges.\\
\textit{The Immersive Frontier:}At its core, spatial computing provides an unparalleled opportunity to transcend the boundaries of traditional user experiences. By seamlessly combining the real and virtual worlds, it opens the door to immersive environments that go beyond entertainment, influencing sectors such as education, training, and gaming. As industries pursue for novel ways to engage audiences and enhance learning outcomes, the immersive potential of spatial computing evolves as a beacon of innovation.\\
\textit{Collaboration Redefined:} In an increasingly interconnected global landscape, collaboration is essential. Spatial computing steps into this field as a catalyst for collaboration progression. Industries like architecture, engineering design, education, and retail are witnessing a revolution as spatial computing simplifies real-time collaboration in three-dimensional spaces. The ability to work together with complex models across distances not only streamlines workflows but also promotes creativity and innovation.\\
\textit{Augmented Efficiency:} Maintenance and repairs in industrial settings have long been afflicted by challenges related to downtime and efficiency. Technicians equipped with spatial computing tools, particularly AR can now overlay digital information onto physical machinery, transforming maintenance processes. This transformative approach reduces downtime, improves accuracy, and sets the stage for a new era in industrial efficiency.\\
\textit{Data Unleashed:} The overflow of data characterizing today's landscape demands innovative solutions for understanding and decision-making. Spatial computing rises to the circumstance, providing a three-dimensional canvas for advanced data visualization and analytics. Sectors ranging from finance to urban planning look at spatial computing as a tool to solve complex datasets, transforming information into actionable insights.

This motivates us to explore spatial computing and unveils the thrilling journey at the technology frontier, bridging the virtual and real worlds. This fusion reforms how we interact with reality and offers insights into the future of AR, VR, and more. As digital experiences interlink with daily life, understanding spatial computing's impact across education, healthcare, gaming, and design is invaluable. This attempt not only enriches our collective knowledge but also empowers us to mold the future of technology and human experiences.

\section{Applications of Spatial Computing}

\begin{figure*}[h!]
	\centering
\includegraphics[width=\linewidth]{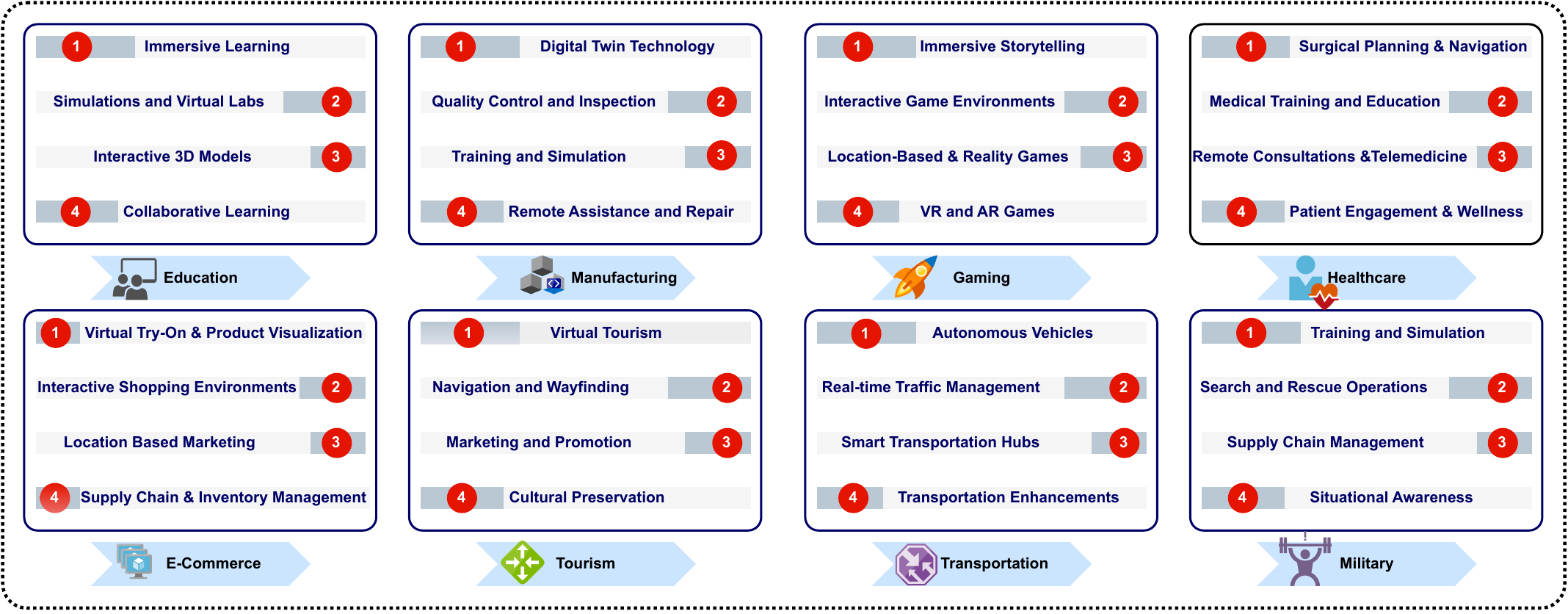}
	\caption{Navigating the Spatial Frontier: Diverse Applications of Spatial Computing.}
	\label{fig:app}
\end{figure*}

%
%
%
In this section, we will explore the numerous applications of spatial computing, investigating how this technology can reshape industries and experiences across sectors such as education, healthcare, the military, e-commerce, manufacturing, transportation, gaming, the architectural sector, transportation, and tourism. Table. \ref{tab:application_table} illustrates the applications of supportive technologies in spatial computing and how they contribute to enhancing the immersive experience in spatial computing. Fig. \ref{fig:app} illustrates the diverse applications of spatial computing across a spectrum of sectors.

\begin{table*}[h!]
\caption{Applications of Supportive technologies of Spatial Computing and its challenges.}
\label{tab:application_table}
\centering
\resizebox{\textwidth}{!}{%
\begin{tabular}{ | m{3.0 cm} | m{6 cm}| m{4 cm} | m{4 cm}|m{4 cm}|}
\hline
\textbf{Technology}    & \textbf{Description}                                                                                                                 & \textbf{Applications}                                              & \textbf{Advantages}                                                     & \textbf{Challenges}                                             \\ \hline
AR \cite{xiong2021augmented} & Integrates digital information and virtual objects into the user's real-world environment, often through mobile devices or headsets & Training and simulation, Gaming, Navigation, Education, Healthcare & Enhances user experience, Real-time interaction with the physical world & Limited field of view, Hardware limitations, Safety concerns    \\ \hline
VR \cite{wohlgenannt2020virtual}   & Immerses users in a completely digital environment, typically through headsets that block out the physical world                    & Gaming, Training, Healthcare, Architecture, Entertainment          & Immersive experiences, Total control over the virtual environment       & Motion sickness, Limited real-world interaction, Isolation      \\ \hline
MR \cite{rokhsaritalemi2020review}    & Blends digital and physical elements, allowing digital objects to interact with the real world                                      & Industrial design, Remote collaboration, Healthcare, Education     & Enhanced user interaction, Real-world integration                       & Complex hardware, Calibration, Limited adoption                 \\ \hline
Spatial Mapping \cite{koziel2008space}       & Uses sensors to create detailed 3D maps of the physical environment, which can be used in AR and MR applications                    & Navigation, Gaming, Architecture, Robotics, Autonomous vehicles    & Accurate environmental understanding, Real-time updates                 & Privacy concerns, Data processing requirements, Cost            \\ \hline
Holography \cite{shi2021towards}             & Creates 3D holographic images that appear in the physical space, often without the need for headsets                                & Medical visualization, Product design, Entertainment, Education    & Realistic visualizations, Minimal device dependency                     & Limited field of view, Complex display technology, Cost         \\ \hline
Spatial Audio \cite{hu2022intuitive}          & Manipulates sound to create immersive audio experiences that match the user's location and orientation                              & Gaming, Virtual tours, Communication, Education                    & Enhanced audio realism, Improved user engagement                        & Audio processing demands, Limited hardware support, Calibration \\ \hline
\end{tabular}
}
\end{table*}.
\subsection{Education}
Modern technologies have transformed the teaching and learning landscape with the introduction of innovative tools and platforms. Learning Management Systems (LMS) have enhanced course management and content delivery, while video conferencing and webinar tools have provided instant communication and collaboration. Moreover, online learning platforms now offer a variety of courses and resources, and mobile learning provides pervasive access to learning content. Adaptive learning systems are used to personalize the learning experience, while collaborative tools promote group work and knowledge sharing. Learning analytics provide useful insights into students' performance. These technologies have empowered teaching-learning by encouraging active participation, personalized instructions, and interactive learning experiences.

However, the integration of these technologies in teaching-learning poses various technological challenges: lack of immersion, hands-on experience in online education, difficulty in understanding and visualizing spatial concepts, limited collaboration and engagement, concerns in accessibility, and issues in conducting evaluations and providing feedback.

\subsubsection*{How Spatial Computing can help}
Spatial computing can provide many advantages in education by utilizing technologies such as VR, AR, MR, and holographic computing. It can be used to develop immersive learning environments, allowing students to experience virtual simulations and historical contexts, resulting in enhanced understanding and engagement. Virtual object manipulation can allow practical experiences and hands-on learning, which may be challenging in real-life environments. Using interactive 3D models and virtual tours, Spatial Computing can improve spatial understanding and visualization. This can help improve the development of abilities for spatial reasoning and cognitive mapping. Collaborative learning can be facilitated in shared virtual spaces, which can enable students to work together on projects and engage in problem-solving. Moreover, spatial computing can address the challenges of accessibility and inclusion by providing various learning styles and individual needs. Assessment and feedback systems can be enhanced using interactive assessments and real-time evaluation. Thus, by integrating spatial computing, educators can create dynamic and engaging learning environments to optimize students' learning outcomes. A few researchers have proposed frameworks/plans for Spatial Computing for education. The research plan presented in \cite{heo2016} aims to create a smart campus using SBD analytics for education, safety, health, and campus management, which involves generating a 3D map, analyzing attendance, collecting student data, and investigating the relationship between trajectory patterns and pedagogical characteristics. This work aimed at predicting student performance, developing proactive student care systems, and providing customized education services. The work in \cite{akers2020} examines a case study of MR spatial computing used in a completely remote undergraduate class. The study used conventional teleconferencing software and MR spatial computing using Magic Leap One headset and a program called Spatial which allows people to interact with each other through virtual avatars. A virtual poster session was conducted using MR in Spatial to display the results of our project. The results depicted that students had a great experience using MR, improved awareness, enhanced collaboration, and felt more connected with other students.

\subsection{Manufacturing} 
Smart manufacturing in Industry 4.0 integrates a variety of enabling technologies to revolutionize conventional manufacturing processes. IoT has become an integral part of smart manufacturing, which facilitates the connection of devices and equipment in order to exchange real-time data and knowledge. Big data analytics and AI provide manufacturers with informed decisions, predict maintenance requirements, and optimize production procedures. Robotics and automation offer streamlined operations to guarantee accuracy and consistency. Additive manufacturing enhances prototype creation and complex production. Advanced sensors along with cloud computing and robust cybersecurity measures, play a vital role in establishing a secure and well-connected manufacturing ecosystem. These technologies have enhanced efficiency, quality, and agility in modern manufacturing industries.

However, the manufacturing industries still face various challenges such as: lack of real-time visualization for identifying inefficiencies and bottlenecks, challenging and error-prone complex assembly tasks, time-consuming training of new employees on intricate machinery, limited adaption to changing demands for customization, challenges in quality control and quality monitoring, limitations in rapid prototyping and testing capabilities, lack of real-time collaboration.

\subsubsection*{How Spatial Computing can help}
Spatial computing has the potential to revolutionize the manufacturing industry by fusing the physical and digital realms. With AR and VR, manufacturers can create digital twins of their products and processes that provide real-time simulations for production lines and assembly processes. Immersive interfaces can allow technicians the ease of remote monitoring and maintenance. Moreover, customer engagement and innovation can be enhanced with collaborative design, customization, and personalized experiences. Additionally, spatial computing can aid supply chain management, quality control, and data visualization, which leads to higher efficiency and flexibility. Even though the adoption would require investments in hardware and software, spatial computing can reward the manufacturing industries with streamlined operations, cost reduction, process automation, and enhanced product quality.

The authors in \cite{fuste2020} propose a user-friendly framework for visually programming robot motion in AR. The proposed system enhances human-robot collaboration in physical spaces and provides ease in creation of various robotic interfaces. The framework is validated through interviews with robotics experts and prototype development using mobile AR to demonstrate its potential for distinct robotic systems. The authors in \cite{guibing2018} emphasizes the need for a secure and dependable manufacturing system. A comprehensive evaluation method is proposed in this work to assess system vulnerabilities which utilizes complex networks and simulations. The approach includes system representation, equipment failure scenarios, vulnerability analysis along with quantification methods. It quantifies vulnerability information with a multidimensional spatial computing method.

\subsection{Gaming and Entertainment} 
In the world of gaming and entertainment, several advanced technologies have come up to enhance player experiences. High-definition graphics is used to create wonderful visuals; AI is employed to offer realistic behaviors of characters; Physics engines are used to simulate realistic interactions; Networking is used for multiplayer and online gaming; Gesture and motion controls are used for immersive gameplay; Advanced audio techniques are used to create rich sound environments; Procedural content generation is used for dynamic gaming worlds; Streaming platforms are used for remote gaming; Machine learning is used for innovative storytelling methods; Cloud computing is used for server support; Biometric sensors are employed for emotional engagement. These modern technologies collectively play a significant role in enhancing the landscape of gaming and entertainment experiences. 

However, there are several limitations of current gaming and entertainment technologies, such as: lack of immersion, limited interactivity, screen dependency leading to issues of eye health, limited social interaction, limited accessibility, static environments, passive entertainment, lack of customization, limited realism and presence, and issues in integrating physical merchandise. 

\subsubsection*{How Spatial Computing can help}
Spatial computing can revolutionize the gaming and entertainment industries with AR, VR, and MR to provide immersive experiences. These technologies can enable players to interact with virtual worlds using gestures and movements. Moreover, they can foster collaborative gameplay and social interactions. Fusing physical and virtual elements can create fascinating and novel experiences. With spatial computing, it would be possible to enhance the creation of user-generated content to build customized virtual environments. On the other side, spatial computing can improve realistic training simulations with the convergence of traditional film-making. Moreover, spatial computing can redefine marketing with AR and VR campaigns and transform user experiences by offering novel perspectives and real-time interactions. Thus, spatial computing can reshape the gaming and entertainment landscape by enhancing content creation, storytelling, social interaction, merchandise integration, and immersive spectator experiences.

A few researchers have proposed frameworks/plans for spatial computing for gaming and entertainment. In \cite{elor2020}, the authors present the concept of Infinite Photogrammetry to showcase its application with modern XR headsets. The Wave Function Collapse algorithm showcased its potential by translating physical environments into virtual spaces. However, the work needs technical refinements with scalability issues unaddressed. The authors in \cite{jot2021} propose an efficient spatial audio rendering system that works with physically accurate environment models and musical soundscapes. This scheme aims at creating immersive experiences in virtual environments where sound sources and acoustic properties can change dynamically with minimum computational load and complexity. The proposed scheme provides ease in manipulating sound source positions, listener positions, room characteristics, and obstacles in order to create realistic audio effects. The editors in \cite{solbiati2020} describe the key benefits of spatial computing over existing methods, such as ease of adaption for younger operators of video games, affordability, and ease in making operative decisions with image-guided techniques. 

\subsection{Healthcare}
The importance of spatial computing in healthcare stems from the transformative potential of this emerging technology. The technologies it encompasses, AR and MR, hold immense promise in revolutionizing healthcare delivery, training, and patient outcomes. By leveraging these technologies, healthcare professionals can enhance surgical planning, interactive and immersive medical training experiences, relieve pain and anxiety during medical procedures, and improve rehabilitation and physical therapy programs.  These technologies can also play a crucial role in remote consultation, telemedicine, medical data visualization, patient education, and medical research. The healthcare industry has experienced exponential growth and transformation with the introduction of cutting-edge technologies. From electronic health records to telemedicine, AI to genomic medicine, and robotics to virtual reality, these technologies adoption has revolutionized healthcare services, patient outcomes, and enhanced research capabilities. The adoption of digitization and connectivity has restructured processes, enhanced efficiency, and eased seamless information transfer. While the healthcare industry has made significant developments in providing improved services through the integration of recent technologies, it acknowledges that there is still room for further improvement in order to provide even better services in terms of accuracy, precision, security, usability, and experience. 

However, even after the adoption of current technologies in healthcare, numerous persistent limitations continue to challenge the industry. These comprise difficulties in data integration and visualization, obtaining precision during surgeries, remote patient care, effective medical training, and patient engagement. In addition, significant concerns still remain, including issues related to mental health support, emergency response, healthcare accessibility, chronic disease management, assistive devices, diagnostic imaging, and patient privacy. 

\subsubsection*{How Spatial Computing can help}
Spatial computing's ability to merge the physical and virtual worlds to offer interactive and immersive experiences can potentially be used in the healthcare industry in several ways. This includes surgical precision, medical training and education, patient engagement, telemedicine, research and data visualisation. It allows surgeons to visualise patient-specific data, like anatomical structures, in a three-dimensional virtual space. It also overlays the critical structures in patient data from the surgeons point of view during the procedure. This augmented visualisation of anatomical structures and potential challenges before surgery could help the surgeons plan the surgical procedure more precisely, considering incision points, approach angles, and potential risks \cite{farhadloo2023spatial}. It can also help in tracking the surgical equipment during the surgery in real-time. Augmented visualisation along with this tracking feature can help in ensuring precise manipulation of instruments, leading to accurate actions during the surgery such as incisions, tissue dissection, and suturing. It also facilitates advanced data visualisation and analysis in medical research. Through the ability to create realistic and virtual environments, it can help researchers study diseases, test new treatments, and analyse complex data. This can accelerate the research perspective of the healthcare industry. It also offers substantial benefits in medical training and education by revolutionising the way students and healthcare professionals learn and practise \cite{sharafy2022future} \cite{pangilinan2019creating}. It helps to create immersive virtual simulations that replicate a realistic view of surgical procedures, patient examinations, or emergency situations. It also helps to have access to a wide range of virtual patients and medical cases with complex pathology's and anatomical variations. It also allows trainees to explore and interact with virtual anatomical structures, enhancing their understanding of the human anatomy of different body systems. It also helps patients understand their health conditions by providing interactive and visual experiences \cite{hoffler2010spatial}. This could lead to better comprehension and increased involvement in their care. It also enables remote collaboration and consultation among healthcare professionals by offering the provision to virtually appear in the patients environment \cite{riva2000virtual}, examine symptoms, and provide guidance from a distance.

It is essential to embrace the power of spatial computing to open up a new era of healthcare by revolutionizing diagnostics, obtaining surgical precision, patient engagement, and remote collaboration to overcome the micro-level challenges of modern healthcare. The authors in \cite{farhadloo2023spatial} have proposed a spatial computing vision model called "Atlas EHR" an alternative way of representing patient histories and biomedical data. This model was proposed to handle the challenges of the current healthcare system. It has been proposed to address challenges such as increase in doctor-patient ratio, lengthy medical histories, emergency treatment demands, and patient variability. The current healthcare system offers a longitudinal view of patient medical history, which is time-consuming, and sometimes healthcare providers demand assistance from others for preliminary analysis. Atlas EHR initially starts with offering an overview of patient health before getting into the spatially anatomical sub-systems and their individual components, or sub-components. This ability of Atlas EHR could potentially help healthcare providers have better insights into complex diseases and also offer tailored treatment plans that can help improve patient outcomes.
\begin{table*}[h!]
\caption{Applications of Spatial Computing.}
\label{tab:app_table}
\resizebox{\textwidth}{!}{%
\centering
\begin{tabular}{|p{2.6 cm}|p{6 cm}|p{5.3cm}|p{4.5 cm}|}
\hline
\textbf{Application   Area} & \textbf{Current   Challenges}                                                                                                                                                                                                                                                                                                & \textbf{Applications of   Spatial computing}                                                                                                                                                                            & \textbf{Supportive   Technologies used}                                                                                               \\ \hline
Healthcare                  & Data   integration and visualization, Precision during surgeries, Remote Patient   Care, Effective Medical Training, Patient Engagement, Diagnostic Imaging                                                                                                                                                                  & Surgical   Navigation, Medical Training and Education, Patient Education, Telemedicine,   Pain Management, and Rehabilitation and Physical Therapy                                                                      & AR,   VR, IoT, Data Analytics, Wearable Sensors, and Telecommunication Technologies                                                   \\ \hline
Gaming                      & Limited   immersion and interaction with the real world, Integrating digital and   physical worlds seamlessly, Lack of immersive audio experiences, difficulty   in sound positioning                                                                                                                                        & Markerless   tracking, real-world interaction, location-based experiences, Immersive,   interactive virtual environments, realistic simulations, Realistic sound   positioning, 3D audio environments                   & AR,   VR, Haptic Feedback, Holographic displays, 3D Modelling, SLAM, Geo-location   Services                                          \\ \hline
Manufacturing               & Lack of real-time visualization for   identifying inefficiencies and bottlenecks,    Challenging and error-prone complex assembly tasks, Time-consuming   training of new employees on intricate machinery, Limited adaption to   changing demands for customization, Challenges in quality control and quality   monitoring & Real-time   collaborative design, 3D modeling and visualization, Mixed Reality in Design,   Robotics with Spatial Awareness, Quality Control and Inspection, Spatial   Computing in Logistics, Assembly Line Assistance & AR,   VR, IoT sensors,Real-time tracking, Data Analytics, 3D Modelling, SLAM,   Geo-location Services, Telecommunication Technologies \\ \hline
Education                   & Lack of immersion, Hands-on experience in   online education, Difficulty in understanding and visualizing spatial   concepts, Limited collaboration and engagement, Concerns in accessibility,   and Issues in conducting evaluations and providing feedback                                                                & Interactive   Learning, Simulated Science Labs, Field Trips in VR, Anatomy and Biology   Education, Language Learning, History Lessons                                                                                  & AR,   VR, 3D Modelling, Simulators and Haptic Feedback, and Telecommunication   Technologies                                          \\ \hline
E-Commerce                  & Product   visualization, Personalized shopping, Supply chain visibility, Returns and   Exchanges                                                                                                                                                                                                                            & AR Product Visualization, Virtual Try-Before-You-Buy, Augmented Reality Shopping Apps,   Geolocation-Based E-Commerce, Augmented Reality Advertising                                              & AR,   VR, Wearable Devices, 3D Modelling, Geo-Location Services                                                                       \\ \hline
Architectural   Design      & Design   Visualization and Feedback, Coordination and Clash Detection, Client   Engagement, User Experience and Functional Design, and Sustainability and   Performance Analysis                                                                                                                                            & Virtual   Design and Visualization, 3D Modeling and Rendering, Augmented Reality for   Architectural Visualization, Virtual Walkthroughs and Tours, Collaborative   Design and Planning                                 & 3D   Modeling and Visualization Software, VR, AR, 3D Scanning and LiDAR Technology                                                    \\ \hline

Tourism & Challenges in providing immersive, personalized experiences and real-time information to travelers, often limited by traditional mediums like static 3D models & Enhancing Engagement, Offering Tailored Experiences, and Improving Sustainability in Tourism & VR, AR, and IoT \\ \hline

Transportation & Virtual Design and Visualization of traffic, 3D Modeling and Rendering of maps, Augmented Reality for  Visualization & Optimizing Routes, Managing Traffic, and Enhancing Navigation. & AR and VR \\ \hline

Military                    & Decision-making   capabilities, and communication efficiencies, adapting to rapidly evolving   threats and ensuring readiness in an ever-challenging landscape demand   continuous improvement                                                                                                                               & Military   Training Simulations, Tactical AR, VR for Combat   Training, Geospatial Analysis and Mapping, Battlefield Visualization                                                          & AR,   VR, Simulators, Vizualization tools and Telecommunication Technologies                                                          \\ \hline
\end{tabular}
}
\end{table*}
\subsection{E-commerce}
E-commerce is the process of purchasing and selling goods and services online. It has become a fundamental and integral part of modern business and consumer culture. It has revolutionised the way customers shop and interact with products. Today's e-commerce has been completely powered by a multitude of technologies. This includes seamless transactions, personalised experiences, and efficient operations. The rapid proliferation and developments of web and mobile technologies serve as the foundation for online retail, offering customers intuitive interfaces and payment gateways. Technologies such as AI and machine learning have enhanced the customer experience while purchasing through chatbots, virtual assistants, and personalised recommendations based on the customers desires. In addition, technology integration into logistics and supply chain management enabled optimised inventory management systems, order fulfilment software, and real-time tracking opportunities. As technologies continue to evolve, they offer more convenience, personalization, and accessibility to consumers worldwide.

Despite the significant technological developments in e-commerce, certain limitations persist that can obstruct the shopping experience for consumers. These limitations include challenges related to product visualization, personalized shopping, supply chain visibility, returns and exchanges, and many others. 

\subsubsection*{How Spatial Computing can help}
Spatial computing has significant scope to contribute to the field of E-Commerce. The supporting technologies of spatial computing, such as virtual and augmented reality, can potentially be used in various ways to enhance E-commerce experiences. Spatial computing can enable virtual trial-on experiences, allowing customers to visualise and try their desired products virtually \cite{cronin2020infinite}. It is considered an innovative application of spatial computing in E-commerce. This virtual trial overcomes the critical challenge of traditional E-commerce, where customers cannot physically try on the products they like \cite{bonetti2018augmented}. Using the virtual avatar created for each customer, they can browse through the catalogue of products such as clothing, accessories, eyewear, and cosmetics. The technology renders the virtual product of the users live video feed. This offers an immersive, realistic way to see how the products look on them. It also offers a way to interact with the virtual product, enables operations such as rotation, zooming, and makes adjustments to get a better understanding of its appearance, fit, and functionality. They can also make comparisons with different products, evaluate their options, and make more informed purchase decisions. It can also help in creating a virtual shopping environment that can provide real-time product visualisation and an immersive trial experience. The creation of a virtual shopping environment \cite{grupac2022virtual} is a method that makes use of spatial computing technologies to replicate the experience of shopping in physical stores within digital space. This virtual shop can allow customers to explore and interact with products in a more engaging and realistic way. 3D models can be used to create a virtual representation of products that will be featured in the virtual store \cite{xi2021shopping}. This includes capturing detailed representations of products, textures, colours, and dimensions. Spatial computing enables customers to interact with products in a virtual store using gestures, voice commands, or controllers. By using AR and VR devices, customers can navigate through the virtual aisles, pick up and examine desired products, and interact with the user interface \cite{pfeiffer2020eye}. Each product in the store can be linked with the information that customers may expect to view, such as product descriptions, specifications, customer reviews, and pricing \cite{elboudali2020customised}. By interacting with the products seamlessly, customers can retrieve relevant information required for shopping. AR allows customers to overlay virtual products onto their realistic environment. This could allow them to visualise how products look in their own space. This virtual store can be made available on multiple platforms, such as smartphones, VR headsets, and AR devices, to cover a wider audience. It also has the potential to transform advertising and marketing with end users. It can offer new and engaging ways to interact with brands and products through spatial computing technologies. Augmented reality and virtual reality can be used to create immersive advertisements for the products of the brands to be kept in the virtual shop. This can allow the user to interact or use technology virtually in their real-time environment. This could foster a strong sense of confidence in the product chosen to purchase and also create an emotional bond with the brand. With these support technologies of spatial computing, brands can showcase product features and functions in an interactive and engaging manner. This can help the brands provide better product demonstrations, highlighting key benefits and uses.

\subsection{Architecture Designing}

Architecture is the art and science of designing and constructing buildings and other physical structures. It comprises the creation of spaces that are functional, aesthetically pleasing, and responsive to the needs of people and the environment. The architecture industry plays a crucial role in shaping the built environment. It includes planning, design, and building construction and structures that serve various purposes in society, from residential and commercial spaces to public infrastructure and cultural landmarks. Architects are responsible for conceptualising creative designs that complement functionality, aesthetics, and sustainability while considering the needs and desires of clients. Advancements in technology have had a tremendous impact on the architecture industry, developing and upgrading traditional design processes and enhancing several aspects of architectural practise. With the introduction of digital technologies, architects now have the possibility of accessing powerful tools and software that can streamline their workflows, improve visualisation, and simplify collaboration. Technologies such as VR, AR, computational design, and 3D printing are some of the technologies that have transformed the architecture industry. These technological advancements allow architects to create highly detailed and realistic 3D visualisations. This provides clients with a clear view and an immersive experience of their proposed designs. VR and AR allow for interactive walkthroughs and contextual understanding, enabling better communication and collaboration throughout the project.

Despite technological developments, the architectural industry struggles with challenges like design visualization and feedback, coordination and clash detection, client engagement, user experience and functional design, and sustainability and performance analysis. Traditional methods can obstruct design communication, and create way to feedback difficulties due to time constraints and budget limitations. Coordination challenges evolve from complex projects, and client engagement remains vital for project success. Designing spaces that provide diverse needs and incorporate sustainable practices continues to pose challenges. 
 
The challenges listed above create a massive demand for technologies in the architecture industry. This refers to the increasing demand for digital tools and solutions that improve design visualisation, streamline workflows, support sustainable initiatives, and smooth collaboration. Architects look for technologies such as VR, AR, and energy-efficient solutions to enhance efficiency and innovation in architectural practise.

\subsubsection*{How Spatial Computing can help}
Spatial computing support technologies like VR and AR provide architects and clients with the ability to step into a digital representation of a construction site or building. With VR, architects can generate fully immersive virtual environments, enabling users to walk through the design, inspect and investigate details, and get a feel of the scale and proportion of a building. AR, on the other hand, overlays digital information in the real-world environment, allowing architects to visualise proposed designs in the context of the real-time physical site. These immersive visualisations provide clients with a realistic experience of the design, allowing for better communication and understanding.

Shapespark \cite{Shapespark} is a recent VR model developed to be used in architectural industry. It is a cloud-based platform that permits users to create and share VR models of their designs that can be visualized in a web browser. It is a visualization platform personalized for architectural and interior design projects. It offers user-friendly, web-based accessibility and instant, high-quality rendering. Architects, designers, and developers make use of this tool to create immersive, interactive 3D walkthroughs of architectural spaces, improving collaboration and design communication. As shapespark supports VR and AR experiences, enables asset import from various design software, and claims customization options to suit specific project requirements. This makes it a precious tool for conveying architectural concepts with realism and interactivity.

\subsection{Tourism} 
Tourism is a multifaceted industry, ranging from sightseeing and leisure travel to business trips and cultural experiences. As one of the largest and most profitable sectors globally, tourism has continually evolved to adapt to consumer needs and emerging technologies \cite{sharpley2014tourism}. Despite significant advances in the industry, various challenges still exist, such as enhancing the tourist experience through real-time information, personalized recommendations, and improved navigation capabilities \cite{gretzel2015smart}. For example, traditional 2D renderings and static 3D models can limit client understanding, the tourism sector also faces challenges in conveying the complete essence of a tourist destination through conventional mediums like photos or basic 3D models \cite{gulmurodov20213d}. Issues of scalability, lack of real-time updates, and the inability to offer personalized experiences based on real-time location plague the industry \cite{zhou2015detecting}. Effective engagement with tourists, both before and during their travels, is another challenge. Tourists need real-time information, which can be particularly critical during peak seasons or events, to optimize their experiences. Sustainability and reducing environmental impact have also become crucial in modern tourism.

\textit{How Spatial Computing can help:}

Spatial computing \cite{shekhar2015spatial}, encompassing technologies like VR, AR, and IoT, offers transformative potential to address these challenges. Spatial computing technologies provide a more interactive and engaging way for tourists to plan and experience their trips \cite{bhatt2015spatial}. 
Platforms like Google Earth VR \cite{chen2020google} offer a never-before-seen kind of 'try before you buy' experience for tourists. Users can virtually visit different locations to decide whether they would like to visit in real life, offering a significant advantage for tourism marketers. VR also has the potential to be used in museums and historical sites to offer a more interactive and educational experience, such as in the British Museum's interactive VR experience, which provides a more engaging way to view ancient artifacts.
AR applications can offer real-time, location-based information to tourists. Apps like Wikitude \cite{bourhim2021augmented} provide context-sensitive, real-time overlays of information about historical sites, restaurants, and other places of interest, directly onto the smartphone screens of tourists as they explore. These overlays can include text, photos, and interactive 3D models, offering a rich, informative experience.
IoT has a wide range of applications in tourism, especially in smart cities. From real-time traffic monitoring \cite{wan2022edge} to environmental data collection \cite{samir2019uav}, these technologies can offer tourists detailed, real-time information to enhance their travel experiences. For example, smart parking \cite{al2019smart} solutions can help tourists find available parking spaces in real time, making city navigation easier.
Some researchers have effectively applied spatial computing technologies in the tourism sector for strategic planning and emergency management. In the Western Cape Province, \cite{wan2022edge} uses Geographical Information System (GIS) and Multiple Criteria Evaluation for gap analysis to identify areas of need and opportunity in tourism services and resource allocation. In emergency management, \cite{wang2014visual} analyzes trajectory datasets and topological higher-order information through visual analytics to make critical decisions, such as identifying alternate evacuation centers. These applications provide valuable insights for both long-term planning and real-time decision-making in the tourism industry.

Spatial computing is well-poised to address several existing challenges in the tourism industry, which has the potential to significantly improve tourist experiences through enhanced engagement, providing real-time information, and offering immersive experiences. It promises to offer solutions that are not only more effective but also more sustainable, marking it as one of the key technologies that could shape the future of tourism. Moreover, by bridging the gap between the digital and physical world, spatial computing offers new avenues to enrich the tourist experience, from planning stages to the trip itself, bringing about a new era in the industry.

\subsection{Transportation} 
Transportation is a major sector that affects revenue growth, social equity, urban development, and sustainable development of the globe. Transportation is not only about moving things from one place to another place, it helps in the growth of cities, increases international trade, and impacts the daily lives of many individuals all over the globe. Transportation has advanced extensively over the years, due to technological advancements, sustainability issues, and the changing needs of globalization. These advancements impact all modes of transportation like land, air, and sea \cite{krishnan2016traffic}.
The recent development of electric vehicles (EVs)  had a significant impact, providing advantages such as reduced pollution and inexpensive maintenance \cite{pramuanjaroenkij2023fuel}. AV, operating using advanced sensors and AI, have the potential to significantly improve collision prevention and travel efficiency \cite{hakak2022autonomous}. Furthermore, many cities are now implementing smart traffic management systems. These technologies, which are powered by AI and IoT, help in efficient traffic management, reducing road congestion and changing traffic lights based on what's happening on the roads at all times \cite{pandya2023federated}. As the world's population increases the transport area faces new challenges in satisfying increasing requirements and minimizing environmental concerns. Achieving a balance among these requirements, innovations in technology, and sustainable growth will be critical in the future. There are significant challenges in the transportation sector \cite{dai2023transport}. One major issue is that transport causes a lot of pollution, particularly greenhouse gas emissions, which must be addressed because many people need to travel. As cities grow in population, roads become increasingly congested, increasing pollution, consuming more petrol, and increasing transportation delays \cite{hossain2023narrowing}. Furthermore, many of the transport systems are old and in poor condition, which can be unsafe and cause problems. Integrating new technology, such as autonomous vehicles, into existing systems is challenging and expensive. Spatial computing, which uses the latest technologies, plays an important role in addressing several issues in the transportation sector \cite{turienzo2023logistics}. These technological advancements improve transportation system performance significantly and create the way for environmentally friendly and practical transportation solutions that benefit everyone \cite{tian2023survey}.

\textit{How Spatial Computing can help:}
Transportation is considered a major source of pollution, mainly in terms of greenhouse gas emissions. Spatial computing plays an important role in decreasing this impact by optimizing travel routes, identifying the shortest and most efficient routes, and lowering fuel consumption and carbon emissions. In EVs, spatial computing helps in energy management and finding nearby charging stations \cite{liu2023distributional}. Spatial computing improves GPS navigation by making navigation more accurate and efficient using AR and VR technologies \cite{rolls2023hippocampal}. BMW recently started using AR Head-Up Displays in its automobiles. Using AR, spatial computing displays driving directions live onto the windscreen. It improves the visibility of the driver's position by displaying directional instructions on the windscreen. Spatial computing enhances the driving experience and also improves safety because drivers can keep their focus on the road when receiving navigational assistance \cite{Morozova_2019}. Similarly, Google Maps added an AR feature called Live View, which is an interesting application for spatial computing. This feature enables the smartphone's camera to capture the actual surroundings and then display navigational details such as indicators and location on the live camera display. This method helps users to determine which way to turn as well as how far they are from their next turn without constantly looking down at their phone, resulting in an interactive navigation experience \cite{Google}. Spatial computing improves traffic management by continuously analyzing and predicting traffic patterns, road conditions, and weather forecasts in the city. Spatial computing assists in providing different paths based on traffic signals to avoid densely populated areas, as well as minimizing traffic delays and improving travel times. Spatial computing significantly improves AV technology by enhancing user safety through the use of sensors and computer vision, improving navigation by precisely navigating the environment and finding optimal routes, enabling interaction between AVs and other travelers. These advancements make AVs a more effective and safe mode of transportation \cite{Av}.

\subsection{Military}
Military and defense are foundational pillars of a nation's security and sovereignty, vital for safeguarding its interests and confirming the safety of its citizens. In an ever-evolving global landscape, the military has not persisted static but has adapted to incorporate cutting-edge technologies to improve its capabilities and effectiveness. The modern military is increasingly characterized by its support on advanced technological innovations, such as AI, cyber warfare, Unmanned Aerial Vehicle (UAV), and sophisticated communication systems. These developments are reforming the nature of warfare, allowing faster decision-making, more precise targeting, and greater overall efficiency. This incorporation of technology into military operations reflects the current drive to stay ahead of emerging threats and maintain a strategic advantage in an increasingly complex world. In this context, it becomes vital to explore how the military harnesses technology to strengthen its defense capabilities and adapts to the ever-changing challenges of the 21st century.

Despite significant developments in technology adoption within the military sector, a number of difficult challenges continue to persist, demanding innovative solutions \cite{karlsson2015challenges}. The dynamic and multifaceted nature of modern warfare scenarios presents a current need for improved situational awareness, decision-making capabilities, and communication efficiencies. The complete volume of data and complexity of information can shatter soldiers and commanders, obstructing their ability to make quick, informed decisions. Meanwhile, continuous coordination among units and effective supply chain management remain crucial for mission success. Additionally, adapting to rapidly evolving threats and ensuring readiness in an ever-challenging landscape demand continuous improvement. These challenges are not only persistent but also increasingly complex.

\subsubsection*{How Spatial Computing can help}
Spatial computing, often associated with AR and MR technologies, holds enormous potential for the military sector due to its unique capability to blend the physical world with digital information in real-time. Spatial computing devices can show real-time sensor data, tactical maps, and other relevant information directly into the field of view of soldiers and commanders. This improved view of the terrain helps them gain a deeper understanding of their surroundings, recognize potential threats, and adapt their tactics accordingly \cite{livingston2011user}. Spatial computing can also be used to provide a comprehensive and up-to-date display of information, permitting military leaders to make better decisions quickly. They can access data such as enemy locations, friendly troop locations, weather situations, and live intelligence feeds, all in a contextual and easily understandable manner. Real-time communication and information sharing are essential demands of military operations. Spatial computing can also help to offer secure, real-time communication and information sharing among military units. Soldiers deployed for military operations in different locations can access the same digital battlefield map and share crucial information instantly. This progresses to more efficient coordination and faster response times during operations.
Simulation and training are vital in the military as they provide realistic, risk-free environments for soldiers to improve their skills, practice decision-making, and get ready for diverse, high-stakes scenarios, ultimately confirming readiness and mission success. With spatial computing, military training can be offered to soldiers at a whole new level. Soldiers can practice in highly realistic augmented environments that replicate the conditions they face on the battlefield. This immersive training through spatial computing helps them develop skills, evaluate different scenarios, and prepare for a variety of mission challenges. Spatial computing can also be used to provide enhanced logistics and supply chain management operations involved in military operations. Spatial computing can be used in aid in asset tracking, inventory control, and optimized routing, confirming that resources are available when it is required. This help to streamline the logistical aspect of military operations, reducing delays and inefficiencies. Spatial computing can also be used to assist in scenario planning, allowing military strategists to create and test different situations in a virtual environment. It can also play a significant role in remote military operations. It improves the control of unmanned systems, drones, and autonomous vehicles by offering more precise navigation, data collection, and surveillance capabilities. This can ultimately increase the effectiveness of these remote assets.

The authors in \cite{Chen_2020} proposed a framework for the command and control system of air defense forces, which makes use of augmented reality technology with Microsoft HoloLens hardware as its foundation. It is a significant advancement in military technology as It offers a dynamic holographic display of battlefield information, improving the commander's ability to perceive and efficiently manage the situation in real-time. By allowing commanders to use gestures and voice commands, as well as arrange simulated military units and interact with weapon systems, the system reduces decision-making time and simplifies collaborative operations. This system not only improves operational efficiency and reliability but also serves as an open platform for operational planning, deduction, simulation, and evaluation. 

\subsection{Summary}
Spatial computing, a transformative technological paradigm, claims a broad spectrum of applications spanning multiple sectors. In education, it aids immersive learning experiences through augmented reality, improving student engagement and comprehension. Healthcare makes use of spatial computing for innovative diagnostic and treatment solutions, enhancing patient care. Gaming and entertainment are transformed by offering captivating, interactive experiences. The architectural sector utilises it for customized design and client engagement, offering advanced visualization and decision-making. In the military domain, it allows advanced training simulations, data visualization, and battlefield planning, improving preparedness and decision-making. E-commerce ties together with spatial computing to create immersive shopping experiences, permitting customers to virtually interact with products before purchasing. In manufacturing, it enhances production processes by providing real-time data visualization, allowing increased efficiency and reduced downtime. Within transportation, spatial computing helps in route planning and autonomous vehicle navigation, resulting in safer and more efficient travel. Furthermore, tourism benefits from AR applications that enrich sightseeing experiences by covering information on historical landmarks and providing interactive guides. However, spatial computing faces challenges such as expensive hardware and software requirements, the design of intitutive and natural UI to interact, content creation and development, privacy concerns in data collection, and security vulnerabilities, as well as social and ethical issues. Understanding these applications and navigating the associated challenges is vital to harnessing the full potential of spatial computing across diverse industries.

\begin{table*}[h!]
\caption{Spatial Computing Projects: Dive into Descriptions, Challenges, and Applications.}
\label{tab:projects}
\centering
\resizebox{\textwidth}{!}{%
\begin{tabular}{ | l |m{2.5 cm} | m{4 cm}| m{3.5 cm} | m{3.5 cm}|m{2.7 cm}|}
\hline
\multicolumn{1}{|l|}{\textbf{Sl.No}} & \textbf{Name of the Project} & \textbf{Description}                                                                                                                   & \textbf{Applications}                                                                                                                   & \textbf{Advantages}                                                                                                          & Challenges                                                               \\ \hline
1.                                   & Neuralangelo                 & Framework designed for 3D surface reconstruction from RGB video recordings                                                            & 3D modeling, Virtual reality and augmented reality, Robotics, Art and entertainment                                                     & Achieves accurate 3D surface reconstruction, high-fidelity models, versatile in capturing both small and large-scale scenes & Still in development                                                    \\ \hline
2.                                   & ProjectDR                    & AR software platform that enables diagnostic images to be projected onto the skin surface of a patient            & Improve the accuracy and efficiency of diagnosis, educate patients about their condition, collaborate between healthcare professionals & Corrects for skin distortion, offers real-time projection, potential to reduce healthcare costs                            & Still in the early stages of development                                \\ \hline
3.                                   & Polycam                      & AR app designed for 3D scanning using LiDAR                                                              & 3D modeling, Architectural visualization, Industrial design                                                                             & Utilizes modern device capabilities for high-quality scans, offers real-time 3D model rendering                             & Still in its early stages of development                              \\ \hline
4.                                   & Project Florence             & An artistic project that explores the possibility of communicating with plants through spatial computing using natural language       & Artistic representation, Scientific Exploration                                                                                         & Two-way communication between plants and humans, integration of various scientific fields                                   & Concept is largely artistic and speculative                             \\ \hline
5.                                   & IKEA Kreativ                 & AI-powered app that allows users to design their own rooms with IKEA products using VR and spatial computing                          & Interior design, Furniture shopping                                                                                                     & Realistic 3D visualization, seamless integration with online shopping                                                       & Reliance on accurate room scanning, user interface complexities         \\ \hline
6.                                   & Magic Leap                   & Florida-based company producing a mixed-reality headset known as Magic Leap 2 (ML2)                                                   & Professional applications like medicine and engineering, enterprise market                                                             & Expanded field of view, lightweight design, immersive AR experience                                                         & Physical comfort, weight distribution, competition in the market        \\ \hline
7.                                   & Apple Vision Pro             & A device that seamlessly integrates digital content with the users physical environment using advanced cameras and spatial computing. & Diverse range of applications due to its integration of the physical and digital worlds                                                & High-definition cameras, intuitive user interface, real-time rendering                                                      & Hardware limitations, competition with other similar devices            \\ \hline
8.                                       & Microsoft HoloLens           & A mixed-reality device that uses waveguide optics to merge real-world vision with computer graphics                                   & Medical, Military, Enterprise solutions                                                                                                & Pioneering in the field of spatial computing, wide range of applications, significant military contract                     & Field of view limitations, reliance on certain tech architectures, cost \\ \hline
9.                                   & Meta Quest Pro               & A VR headset with enhanced spatial computing capabilities                                                                             & Gaming, Mixed Reality experiences                                                                                                      & Enhanced pixel count for better visuals, improved hand tracking, immersion features                                         & Absence of a hardware-level depth sensor, ongoing development           \\ \hline
\end{tabular}
}
\end{table*}
\section {Projects}
In this section, we will explore a wide array of innovative spatial computing projects. These projects showcase the practical applications and creative possibilities of this emerging technology. Table \ref{tab:projects} is a comprehensive overview of spatial computing projects and their respective real-world applications.
\subsubsection{Neuralangelo}
Neuralangelo \cite{2306.03092} is a framework designed to achieve accurate 3D surface reconstruction from RGB video recordings, ensuring exceptional quality. Through the utilization of common mobile devices, this framework empowers individuals to generate precise digital replicas of various scenes, both on a smaller scale, focusing on objects, and on a larger scale, encompassing extensive real-world environments. The outcome is characterized by intricate 3D geometries and a level of fidelity that is remarkably high. 
Neuralangelo employs a methodology to densely reconstruct scene structures through the analysis of multi-view images. This process involves the sampling of 3D coordinates along the various camera view directions. Subsequently, a multi-resolution hash encoding technique is applied to encode these spatial positions. These encoded attributes serve as inputs for two specific neural networks: the Signed Distance Function (SDF) Multi-Layer Perceptron (MLP) and the Color MLP. The SDF MLP works to interpret the encoded features and generate a volumetric representation based on SDF values, while the Color MLP handles the amalgamation of images through SDF-driven volume rendering. Some possible applications of Neuralangelo:

\textit{3D modeling}: Neuralangelo can be used to create high-fidelity 3D models of objects and environments. This could be used for a variety of purposes, such as product design, architectural visualization, and historical preservation.

\textit{Virtual reality and augmented reality}: Neuralangelo can be used to create virtual reality and augmented reality experiences that are more realistic and immersive. This could be used for games, training simulations, and educational applications.

\textit{Robotics}: Neuralangelo can be used to create 3D models of objects and environments that can be used by robots for navigation and manipulation. This could be used for industrial automation, search and rescue, and healthcare applications.

\textit{Art and entertainment}: Neuralangelo can be used to create new forms of art and entertainment that are more interactive and immersive. This could be used for games, movies, and music videos.

\subsubsection{ProjectDR}
ProjectDr \cite{projectdr_link} is an AR software platform that enables diagnostic images to be projected onto the skin surface of a patient. This technology also corrects the image to account for distortion caused by skin contours as well as the size and shape of the individual. ProjectDr was developed by a team of researchers at the University of Alberta's Faculty of Rehabilitation Medicine.

The foundation of ProjectDr lies in the adept utilization of spatial computing techniques. To begin, it employs the prowess of computer vision to meticulously track the patient's bodily features and pinpoint the precise skin area where the image should be projected. Subsequently, it employs intricate 3D modeling \cite{rehab_magazine} to construct a digital replica of the patient's body. This virtual model serves the crucial purpose of rectifying any image distortions arising from the skin's inherent topography. Finally, the magic of AR rendering comes into play, allowing us to seamlessly project the diagnostic image onto the patient's skin in real-time.

ProjectDr is a significant advancement in the field of spatial computing. It demonstrates the potential of AR to be used for medical applications. ProjectDr could be used to improve the accuracy and efficiency of diagnosis, to educate patients about their condition, and to collaborate between healthcare professionals. Additionally, ProjectDr could reduce the cost of healthcare by eliminating the need for patients to travel to radiology labs for diagnostic imaging.

ProjectDr is still in the early stages of development, but it has the potential to make a major impact on healthcare. It is a promising technology that is worth watching in the years to come.

\subsubsection{Polycam}
Polycam \cite{polycam_website} is an advanced AR app designed for 3D scanning. Created by the team at Occipital, Inc., this app utilizes the LiDAR scanner found in iPhones and iPads to generate high-quality 3D models of various objects and environments.

The core of Polycam's functionality lies in its effective use of spatial computing techniques. To start, the app employs computer vision to monitor the user's movements and identify the object or setting being scanned. Next, the LiDAR scanner measures the distance between the user's device and the target. This information is then used to create a "point cloud," which essentially gathers points in 3D space. Subsequently, 3D modeling is used to shape this point cloud into a "mesh," a network of triangles representing the object's surface. Finally, AR rendering brings it all together, allowing the 3D model to be viewed on the user's device in real time.

Polycam signifies a notable advancement in spatial computing. It serves as a prime example of how AR can have multiple applications, including 3D modeling, architectural visualization, and industrial design. Moreover, the app has the potential to create augmented reality experiences that enable users to interact with virtual objects in real-world environments.

Although Polycam is still in its early stages of development, it possesses the potential to greatly influence our interaction with the world around us. This promising technology warrants attention as it continues to evolve in the future.

\subsubsection{Project Florence}
Project Florence is an innovative and artistic project that explores the possibility of communicating with plants through spatial computing using natural language. The project aims to create a plant-human interface experience that is based on scientific analysis of the plant and its environment. The project uses a combination of biology, natural language research, design, and engineering to enable people to converse with a plant by translating their text sentiment into a light frequency the plant can recognize and respond to.

The project consists of a custom-built device that connects to a plant and monitors its bioelectrical signals, soil moisture, temperature, and light levels. The device also has a light source that can emit different colors and intensities of light to the plant. The device is connected to a cloud service that analyzes the plant's signals and environmental data using machine learning algorithms. The cloud service also receives text input from a human user and converts it into a sentiment score using natural language processing techniques. The sentiment score is then mapped to a light frequency that corresponds to the plant's optimal photosynthesis range. The device then emits the light frequency to the plant as a form of feedback.

The project also allows the plant to respond to the human user by generating text output based on the plant's signals and data. The cloud service uses natural language generation techniques to create sentences that reflect the plant's state and preferences. The sentences are then displayed on a screen or spoken by a voice synthesizer. The project thus creates a two-way conversational experience between the plant and the human user.

Project Florence is an artistic representation of a plant-human interface experience that challenges the conventional notions of communication and interaction with the natural world. The project invites people to imagine what plants might say if they could talk and how they might respond to human emotions and intentions. The project also raises questions about the ethical and social implications of creating such interfaces and how they might affect our relationship with nature. Project Florence is a creative exploration of the potential of natural language as a medium for connecting with other living beings \cite{steiner2017project}.

\subsubsection{IKEA Kreativ}
IKEA Kreativ is an AI-powered app that allows users to design their own rooms with IKEA products. The app uses a combination of virtual reality, spatial computing, machine learning, and 3D-mixed-reality technologies to create realistic and interactive 3D replicas of users' spaces. Users can scan their rooms with the IKEA app and erase unwanted items. They can also add new furniture and accessories from the IKEA catalog or explore pre-furnished rooms for inspiration. The app helps users visualize how IKEA products would look like in their spaces and homes before they buy them. It also allows users to save, share, and shop their design ideas online or in the app. According to designboom magazine, the app invites customers to explore products in 50 inspirational 3D showrooms where they can quickly swap, rotate, stack, and hang different objects. Through its LiDAR-enabled IKEA Kreative Scene Scanner, customers can also create lifelike virtual replicas of their spaces in accurate dimensions and perspectives. After designing their ideal space, customers can seamlessly add products directly to their cart and share designs with friends and family \cite{ikeakreativ2023}.

\subsubsection{Magic Leap}
Magic Leap is a Florida-based company working on "mixed-reality" technology that allows you to see virtual objects in the real world. 

Magic Leap's second-generation AR headset, the Magic Leap 2 (ML2), represents a significant advancement in augmented reality technology. One of the key improvements is its expanded field of view (FOV), which has been increased to an impressive 70 degrees. This allows users to experience a more immersive and natural AR environment, with the ability to perceive taller objects without needing to move their head up and down. This FOV enhancement was achieved through a custom architectural approach that combines Liquid Crystal on Silicon (LCoS) technology, LED RGB light modules, and intricate systems of concentrators and polarizers \cite{magicleap2022}.

To address the issue of weight and comfort, Magic Leap reduced the volume of the ML2 headset by over half compared to its predecessor. This resulted in a 20\% weight reduction, making the device feel more akin to wearing traditional eyeglasses. The physical design of the headset has been improved as well, featuring flatter lenses and slimmer arms for a more appealing and ergonomic appearance.

A standout feature of the Magic Leap 2 is its unique global dimmer module. This technology enables the headset to dynamically adjust the level of light entering the user's field of view. It can darken the real-world environment, effectively eliminating distractions and enhancing the visibility and vibrancy of virtual objects. This capability allows users to experience a more VR-like immersion while retaining the ability to interact with their surroundings.

The ML2's sophisticated projection technology combines LCoS and LED RGB light modules, allowing it to produce high-quality, vibrant virtual images. Developers can leverage this technology to create compelling and lifelike AR experiences, further enhancing the sense of presence. Additionally, the headset's new Android-based operating system simplifies the development process and offers a more accessible platform for creators.

Magic Leap's focus on the enterprise market aligns with the headset's technical capabilities. It targets professionals like doctors and engineers who can benefit from AR assistance during tasks such as surgery or working with complex machinery. This strategic shift, coupled with technological advancements, positions Magic Leap as a notable player in the AR landscape, poised to compete with the likes of Microsoft's HoloLens and Meta's VR offerings \cite{uploadvr2022}.

\subsubsection{Apple Vision Pro}
The Apple Vision Pro \cite{apple_vision_pro} signifies a paradigm shift in the realm of spatial computing, seamlessly integrating digital content with the user's physical environment. This integration fosters a sense of presence and connectivity, thereby enhancing the user experience. The device offers an expansive platform for applications, breaking free from the limitations of traditional displays and introducing a three-dimensional user interface. This interface is controlled through intuitive inputs such as eye movements, hand gestures, and voice commands.

The Vision Pro is equipped with a pair of high-definition cameras that transmit over one billion pixels per second to the device's displays. This high pixel density ensures a clear and accurate depiction of the user's surroundings. The system also supports precise tracking of head and hand movements, real-time three-dimensional mapping, and recognition of hand gestures from various positions.

The device operates on visionOS \cite{apple_visionos}, an innovative spatial operating system that enables users to interact with digital content as if it were physically present in their environment. The Vision Pro's design incorporates a unique dual-chip system, powered by the potent M2 chip and the new R1 chip. The M2 chip efficiently runs visionOS, executes advanced computer vision algorithms, and delivers stunning graphics. Concurrently, the R1 chip processes input from the cameras, sensors, and microphones, streaming images to the displays within 12 milliseconds, thereby providing a virtually lag-free, real-time view of the world.

The Vision Pro's design also includes a three-dimensionally formed laminated glass that seamlessly integrates with an aluminum alloy frame, designed to contour to the user's face. The device's ultra-high-resolution display system, consisting of 23 million pixels across dual displays, provides a pixel density that surpasses that of a 4K television for each eye.

Additional features such as the LiDAR Scanner and TrueDepth camera work in unison to create a fused 3D map of the user's surroundings. This enables the Vision Pro to render digital content accurately within the user's space. Infrared flood illuminators also work with the external sensors to enhance hand tracking performance in low-light conditions.

The Vision Pro, a testament to Apple's extensive experience in designing high-performance, mobile, and wearable devices, merges advanced technology with a sleek, compact design to deliver an unparalleled user experience.

\subsubsection{Microsoft HoloLens}
The Microsoft HoloLens \cite{microsoft2023}, a groundbreaking device in the field of spatial computing, overcame numerous developmental hurdles. Despite these, it played a significant role in shaping the future trajectory of the industry. The HoloLens employs ``waveguide" optics, a flat optical system that fuses real-world vision with computer graphics. Utilizing a diffraction grating, a type of Diffractive Optical Element (DOE), the waveguide bends or ``incouples" the light so it undergoes Total Internal Reflection (TIR) within the glass waveguide. The light then rebounds off the flat surfaces within the glass, encounters a triangular ``fold zone," and is directed towards an ``exit zone" DOE. Here, the angle of the light is reduced so it no longer experiences TIR and can exit the glass towards the viewer's eye. Despite being a technological marvel, this intricate optical system has been associated with high manufacturing costs and low yield rates, potentially affecting the affordability of the HoloLens for consumers \cite{ref1}.

One notable limitation of the HoloLens was its Field of View (FoV). The initial generation provided a narrow and square FoV, centered within the user's sightline. Although the second iteration expanded the FoV slightly, it still fell short compared to competitors like the Quest 2. This restrictive FoV could lead to a less immersive experience as the digital overlays did not cover the user's entire visual field.

Further compounding the device's limitations was its reliance on the Intel x86 architecture. Notorious for high energy consumption and heat production, the x86 architecture introduced performance issues in low power mode, negatively impacting the overall user experience.

Despite these challenges, the HoloLens has had a lasting impact on the spatial computing industry. Its initial applications were primarily in scientific and medical domains, with trailblazing engineers developing remote surgical techniques using the HoloLens. This use case demonstrated the device's potential, further reinforced by widespread adoption from corporations like Dell and Nike for various applications.

The HoloLens has also made considerable advancements within the military sector. In 2021, Microsoft secured a contract worth \$21.9 billion with the US Army to supply 120,000 headsets modeled on the HoloLens product line. Spanning a decade, this agreement represents the most significant governmental transaction within the virtual and augmented reality headset industry to date.

The specific model procured under this contract, known as the Integrated Visual Augmentation System (IVAS), integrates various HoloLens-inspired sensors with military-specific features. These include thermal and night vision cameras, visual target acquisition, vital sign monitors, and a concussion detection system. This extensive array of capabilities illustrates the broad applicability and potential of HoloLens technology in diverse fields.

In conclusion, the HoloLens, despite failing to meet initial expectations, has been pivotal in advancing the field of spatial computing. The lessons learned from both its successes and shortcomings would shape the development of future spatial computing devices. Its continued relevance underscores the significance and potential of spatial computing in reshaping the landscape of human-computer interaction.

\subsubsection{Meta Quest Pro}
The Meta Quest Pro \cite{meta2023} signifies a substantial leap in spatial computing capabilities within the VR landscape. Despite its game-centric heritage, this headset incorporates a variety of spatial computing technologies aimed at establishing an immersive and interactive MR experience.

One of the key attributes of the Quest Pro is its Passthrough feature, which utilizes forward-facing cameras for environmental depth measurement. These cameras, aided by 3D reconstruction algorithms, map the physical environment, offering a real-time feed within the headset. However, the absence of a hardware-level depth sensor, as observed in earlier prototypes \cite{metaquest2023}, limits the resolution of this mapping and its subsequent real-world rendering.

Nevertheless, the enhanced pixel count of these cameras has resulted in improved hand tracking capabilities, making controller-free VR interaction more intuitive. This aligns with recent advancements in neural networks, rendering a hardware depth sensor potentially redundant for high-quality hand tracking.

Eye and face tracking are other significant spatial computing elements incorporated in the Quest Pro. These functionalities contribute to heightened realism in VR interactions, enabling detailed expression mirroring in avatars. Despite some limitations in capturing nuanced expressions, these features demonstrate promising implications for future social VR experiences.

The Quest Pro's eye tracking also facilitates foveated rendering \cite{robertson2023}, a technique that optimizes the sharpness of VR displays based on the user's gaze direction. This approach saves on processing resources by only rendering areas of focus in high detail, thereby enhancing overall visual quality, especially given the constraints of the mobile chipset.

The headset also exhibits notable strides in color passthrough, a feature essential for robust MR experiences. However, current limitations in visual fidelity, particularly in varying light conditions, hinder the full potential of this feature.

In summary, the Quest Pro exemplifies an ambitious step forward in spatial computing for VR headsets, exhibiting features that elevate user immersion and interaction within MR environments.

\begin{figure*}[h!]
	\centering
\includegraphics[width=\linewidth]{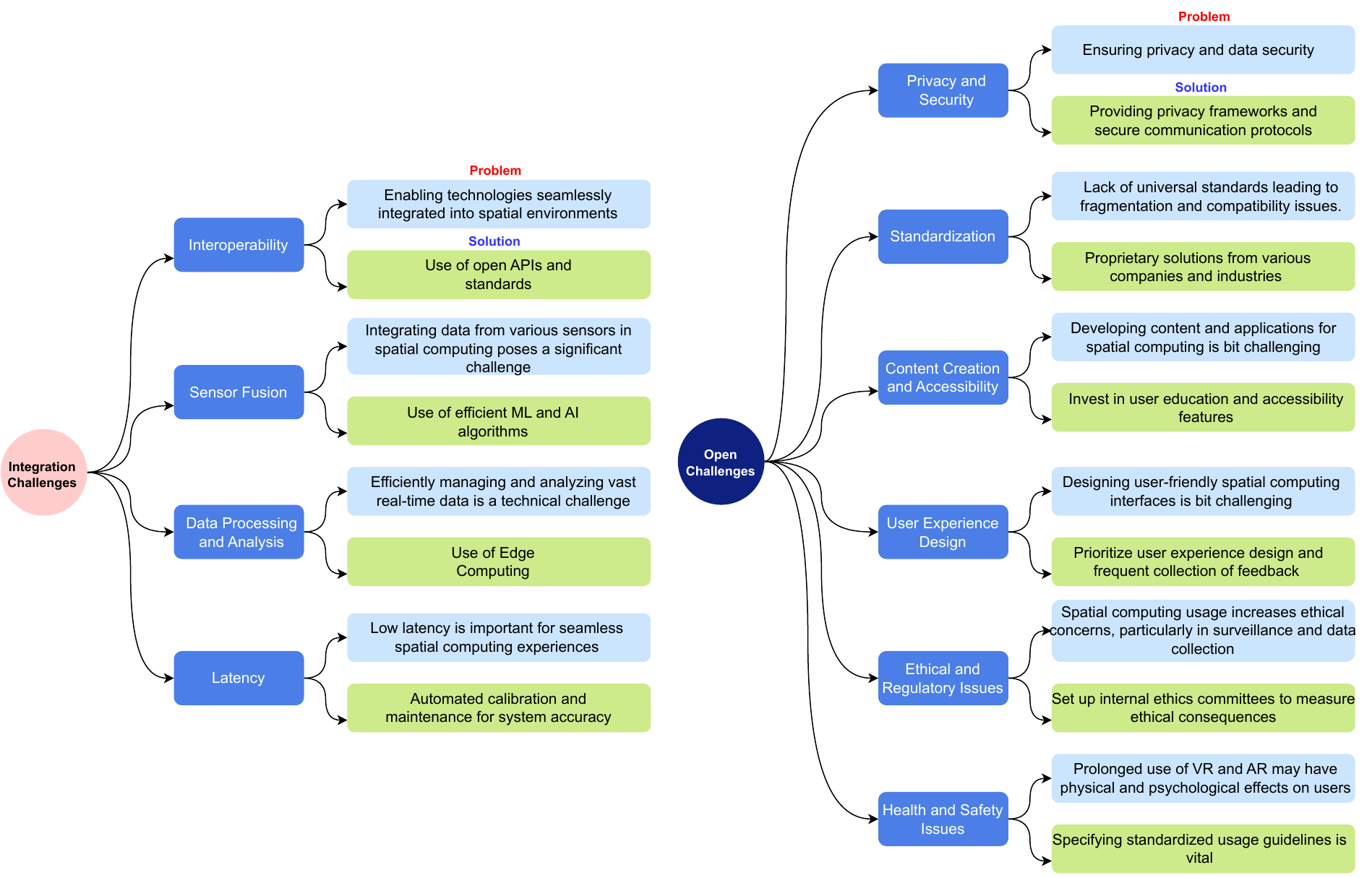}
	\caption{Spatial Computing  Integration and Open Challenges.}
	\label{fig:challenges}
\end{figure*}
\section{Challenges and Future Directions}
%

 In this section, we discuss various challenges posed by spatial computing in the current landscape along with the future directions. 
Fig. \ref{fig:challenges} is a visual representation of the challenges and open issues within the field of spatial computing, complemented by suggested solutions for each. 
 Despite the potential of spatial computing across a range of applications, there remain several challenges to be addressed using some future developments.
 \subsubsection{Interoperability in Spatial Computing}
To achieve seamless user experiences in spatial computing, interoperability is crucial. This is achieved through open APIs and established data standards, enabling communication across diverse technologies. A notable advancement in this area is the integration of MR and robotics, enhancing intuitive interactions between humans and robotic systems \cite{delmerico2022}. The development of open standards and protocols will facilitate cross-platform interoperability, allowing users to seamlessly switch between different spatial computing devices and platforms. This will be essential for enabling widespread adoption of spatial computing and ensuring that users are not locked into a single platform.

\subsubsection{Sensor Fusion and Augmented Reality} 
Sensor Fusion and AR within the realm of spatial computing present intricate challenges that demand meticulous attention. The fusion of data from diverse sensors, such as cameras, accelerometers, and gyroscopes, can lead to complexities in synchronizing and harmonizing information accurately. Discrepancies in sensor accuracy and calibration introduce challenges in achieving precise spatial alignment, impacting the fidelity of AR overlays. Additionally, real-time processing demands for seamless AR experiences require efficient algorithms for sensor data fusion, which can strain computational resources. To address these challenges, a strategic solution involves implementing sensor fusion algorithms with adaptive filtering techniques, such as Kalman filters, to enhance accuracy and synchronization. Leveraging edge computing for on-device processing can alleviate computational burdens, optimizing real-time performance. Furthermore, deploying machine learning models for continuous calibration and dynamic adaptation to sensor variations ensures robustness in spatial computing applications, contributing to a refined and reliable AR experience. \cite{heemsbergen2021}.

\subsubsection{Data Processing and Analysis: A Challenge in Spatial Computing}
One of the significant challenges in spatial computing is the effective processing and analysis of large and complex data sets. This task is crucial for accurately interpreting spatial information and making timely decisions based on it. 

For instance, in epidemiology, spatial computing is employed to manage and analyze health and geographical data. This analysis is vital for tracking disease spread, identifying hotspots, and informing public health strategies. The ACM SIGSPATIAL workshop on Spatial Computing for Epidemiology \cite{acm2021} highlights the importance of sophisticated data processing techniques in understanding and responding to health crises, underscoring the critical role of efficient data handling in spatial computing applications. Addressing the challenge of processing large spatial datasets involves leveraging distributed computing frameworks like Apache Spark or Dask, enabling parallelized computation for efficient analysis. Implementing advanced spatial indexing techniques, such as R-tree or Quadtree structures, optimizes the retrieval of relevant spatial information, enhancing the speed and accuracy of decision-making in spatial computing applications, particularly in fields like epidemiology.

\subsubsection{Latency and Map Rendering for AR}
Low latency is paramount in spatial computing environments, particularly for applications that require real-time interaction and responsiveness. This is essential in AR, where timely processing and rendering of spatial data directly affect user experience and system efficacy. Strategies to reduce latency include optimizing network infrastructures and leveraging edge computing, which brings computational resources closer to where data is generated. A breakthrough in this domain is the development of a platform-independent map rendering framework for mobile AR, as reported by Huang et al. \cite{huang2021}. This innovation enhances the performance and consistency of spatial data visualization across various platforms, which is critical for AR applications that require accurate and quick rendering of complex spatial environments. This advancement also underscores the importance of efficient data processing techniques in reducing latency, thereby improving the overall utility and applicability of AR in spatial computing. The development of high-fidelity 3D maps and accurate localization techniques will allow spatial computing devices to seamlessly understand their surroundings, enabling more realistic and contextual interactions with the physical environment \cite{10.1145/2756547}. This will be crucial for enabling applications such as indoor navigation, augmented reality tourism experiences, and virtual reality simulations that accurately reflect the real world.

\subsubsection{Hardware and Software} Spatial computing experiences usually require high-quality, expensive and resource-intensive hardware such as VR headsets, AR glasses, sensors and others. It is also challenging to manufacture such powerful devices for delivering immersive experiences for widespread acceptance. Moreover, employment of spatial computing  often comes with a significant cost in terms of software development. Thus, factors such as software development and hardware cost can limit accessibility to a wider population, especially in organizations with limited budgets. Mitigating the accessibility challenges in spatial computing involves optimizing hardware through edge computing paradigms, reducing reliance on high-end devices. Utilizing cloud-based rendering and processing can offload resource-intensive tasks, making immersive experiences more accessible without compromising quality. Additionally, adopting open-source development frameworks and cross-platform compatibility in software design minimizes development costs, fostering wider adoption across diverse budget constraints. The increasing availability of powerful edge computing devices and advancements in sensor fusion will enable spatial computing systems to process data and make decisions locally, reducing latency and improving performance. This will be essential for enabling real-time interactions with the physical environment, such as augmented reality overlays that provide real-time information and guidance. One such example is the Humane AIpin \cite{Humane}, integrating Artificial Intelligence with Real Time Spatial Interaction.

\subsubsection{User Interface and Interaction Design}
It is challenging to design user interfaces which are intuitive and natural for a quality immersive user experience. Traditional 2D interfaces may not be effectively translated to 3D space. Thus, designing smooth and non-intrusive interactions along with optimizing user engagement remains a big challenge. The integration of haptic feedback and immersive interfaces will enhance the realism and engagement of spatial computing experiences, allowing users to feel virtual objects and interact with them in a more natural way. This will be essential for enabling applications such as virtual reality training simulations, haptic feedback for virtual assembly tasks, and immersive gaming experiences.


\subsubsection{Privacy and Data Security} 
Spatial computing devices often require extensive data collection to function effectively. This can include capturing real-time video, audio, location data, and even biometric data like eye movements or facial expressions. Storing and managing this vast amount of data securely is a significant challenge. With the ability to track a user's movements, gaze, and interactions in a virtual environment, there's potential for creating detailed behavioral profiles. These profiles can be used for targeted advertising or other purposes, potentially without the user's knowledge or consent. Spatial computing devices, especially those used outdoors like AR glasses, can continuously track a user's location. This raises concerns about stalkers, advertisers, or malicious actors accessing this data. Thus, addressing these challenges requires a combination of technological solutions, robust regulations, and user education. 
As spatial computing becomes more prevalent, it will be crucial to address privacy, security, and ethical concerns related to data collection, user monitoring, and potential biases in AI algorithms. This will require the development of robust privacy-preserving technologies, clear ethical guidelines for spatial computing applications, and public awareness campaigns to educate users about their privacy rights.

\subsubsection{Social and Ethical Considerations} 
Spatial computing, encompassing AR, VR, and IoT, stands at the forefront of technological convergence, yet it navigates an intricate web of ethical considerations. These technologies intersect with data privacy, presenting risks like vulnerability to cyber-attacks and the often absent informed consent for data collection. Additionally, they encounter complexities in algorithmic transparency, where opaque models can result in uninterpretable and potentially biased outcomes, underscoring the imperative for explainable AI systems.

In terms of human-machine interaction, the integration of these systems into daily life raises safety concerns and questions about the cognitive demands placed on users. The potential repurposing of such technologies for surveillance applications further intensifies the urgency for preemptive ethical design principles. It's critical that industry stakeholders set up internal ethics committees to continuously assess and mitigate ethical risks, ensuring that spatial computing advances in a manner that is secure, transparent, and aligned with societal values. This proactive approach to ethics in design and implementation is not just a regulatory safe harbor but a foundational element for sustainable innovation in the spatial computing domain.

\subsubsection{Motion Sickness and Comfort}
The advent of autonomous vehicles and VR systems has expanded the horizons of spatial computing, which encompasses the interaction between humans and technology in both physical and digital spaces. However, a significant challenge that spatial computing faces is motion sickness. This physiological issue is caused by a mismatch between visual, vestibular, and proprioceptive inputs, leading the brain to receive conflicting signals about the body's movement in space. In the context of autonomous vehicles, the issue is exacerbated as passengers may not focus on the horizon or engage with traditional driving controls, increasing sensory conflict.

Quantifiable data from a University of Michigan Transportation Research Institute study suggests that up to 10\% of American adults could experience motion sickness in fully autonomous vehicles \cite{Jones2019}.
 This is a significant proportion when extrapolated to the general population. To address this, designers have proposed mitigative strategies such as installing large, transparent windows, configuring video displays to promote looking straight ahead, and eliminating swivel seats. These measures aim to align the visual and vestibular systems to reduce the likelihood of experiencing motion sickness.

In addition to physical design changes, technology companies like Apple are exploring VR systems specifically designed to mitigate motion sickness in vehicles. Apple's patent application outlines a system that utilizes sensors to monitor indicators such as sweating, pulse rate, and fidgeting to dynamically adjust the VR experience and reduce discomfort. The system also proposes the incorporation of visual anchors and cues that synchronize with the vehicle's real-world movements, offering a more cohesive sensory experience.

While these solutions are promising, they have not yet been tested at scale, and patents do not guarantee that the concept will evolve into a viable product. Moreover, these mitigative strategies need to undergo empirical testing through controlled, double-blind studies across diverse demographic groups to establish their effectiveness conclusively. Therefore, while spatial computing holds great promise, the challenge of motion sickness requires empirical, design-based solutions to ensure the technology's comfort and accessibility.

\subsubsection{Regulatory Hurdles}
Navigating the regulatory landscape for spatial computing presents a multi-faceted challenge, encompassing not just legal but also technical complexities. On one hand, there is a glaring absence of universally accepted standards that define the operational parameters for hardware and software components, leading to fragmented ecosystems that can inhibit interoperability and scalability. Data privacy poses another critical issue; spatial computing inherently collects granular, high-dimensional data about physical environments and user interactions, necessitating robust encryption and anonymization techniques to comply with data protection laws like General Data Protection Regulation (GDPR). Additionally, the technology's capacity for real-time data analysis and augmented reality overlays brings forth ethical quandaries related to surveillance and informed consent, issues that current legislation is ill-equipped to address. The amalgamation of these factors creates a convoluted regulatory environment that necessitates a balanced, data-driven approach to foster innovation while ensuring public safety and ethical compliance. Navigating the regulatory landscape for spatial computing involves addressing the absence of universal standards and privacy concerns. One potential solution is to establish a consortium of legal, technical, and policy experts to develop a comprehensive framework. This framework should include technical standards for interoperability, encryption, and anonymization to comply with data protection laws. Adaptive governance mechanisms can also be integrated to address emerging ethical concerns, promoting a balanced regulatory environment that encourages innovation while prioritizing public safety and ethical compliance.

\subsubsection{Content Curation and Quality Control}
Content creation and accessibility are pivotal aspects in the realm of spatial computing. Crafting content and applications tailored to AR, VR, and IoT is notably challenging due to the diverse user requirements and the intricate nature of immersive environments. A technical addition to the passage could be:

To surmount these challenges, developers must invest in robust content management systems (CMS) tailored for spatial computing, which handle the multifaceted demands of three-dimensional content creation and distribution. Additionally, prioritizing user education is essential to enhance User Experience (UX) design, by integrating interactive tutorials and real-time assistance within the applications. Accessibility features, such as voice navigation and gesture recognition, need to be embedded deeply into system architectures to ensure inclusivity for all user demographics. Furthermore, the adoption of universal design principles ensures that spatial computing applications remain accessible, not only enhancing usability for users with disabilities but also creating a more intuitive interface for the broader user base. Hence, the technical blueprint for quality control in spatial computing must encompass CMS optimization, educational user interfaces, and comprehensive accessibility integration to address the dual challenges of content creation and system accessibility.
\subsubsection{Adoption and User Education}
Adoption and user education present formidable challenges in the proliferation of spatial computing technologies. The technology, while transformative, is often not intuitive for the average user, demanding a steep learning curve that can hinder widespread adoption. The interfaces, gestures, and interaction paradigms are fundamentally different from those of traditional computing systems, necessitating comprehensive educational programs and intuitive design solutions to bridge the knowledge gap. Moreover, the lack of established best practices for user interaction with spatial computing environments complicates the creation of standardized educational materials. In addition, there is the challenge of "digital literacy," where users need to understand not just how to operate within a spatial computing environment, but also how to do so safely and responsibly, given the technology's capacity for data collection and surveillance. These educational imperatives also have to adapt to rapidly evolving hardware and software capabilities, making it difficult to create training materials that remain current. Overall, the dual challenges of user adoption and education necessitate a multi-pronged approach that combines technical innovation with robust educational strategies to ensure that spatial computing can reach its full societal and economic potential \cite{yenduri2023generative}. Making spatial computing devices more affordable and accessible to a wider range of users will be essential for its widespread adoption and societal impact. This will require the development of more affordable hardware, the creation of accessible software and content \cite{halaravr}, and the implementation of policies that promote the availability of spatial computing technologies in underserved communities.

\section{Conclusion}
Spatial computing plays a pivotal role in promoting the seamless integration of the digital and physical realms, hence establishing a seamless interaction between these two domains. This review provides an in-depth evaluation of spatial computing and its enabling technologies. The review additionally highlights the significant impact of the technology on diverse industries, ranging from education to defence. However, the extensive implementation of spatial computing presents significant challenges in the areas of privacy, security, standardisation, ethics, and health and safety issues. Furthermore, the review presents a range of potential solutions to address these challenges. In conclusion, spatial computing represents a substantial technological advancement that significantly transforms engagement with the surrounding environment. However, further investigation is required in the context of spatial computing to effectively integrate the realms of digital and physical environments.

\bibliographystyle{elsarticle-num}
\bibliography{ref}

\end{document}